\documentclass[numberedappendix,letterpaper]{emulateapj}
\usepackage{amstext}
\usepackage{apjfonts}
\usepackage{amsmath}
\usepackage{amssymb}
\usepackage[colorlinks=true,linkcolor=blue,citecolor=blue]{hyperref}
\usepackage{comment}
\usepackage{natbib}

\begin{document}

\newcommand{\acc}{{\bf a}}
\newcommand{\beq}{\begin{equation}}
\newcommand{\eeq}{\end{equation}}

\title{Formation of Tidal Captures and Gravitational Wave Inspirals in Binary-Single Interactions} 
\author{Johan Samsing$^{1,*}$, Morgan MacLeod$^{2,*}$, Enrico Ramirez-Ruiz$^{3}$} 
\altaffiltext{1}{Department of Astrophysical Sciences, Princeton University, Peyton Hall, 4 Ivy Lane, Princeton, NJ 08544, USA}
\altaffiltext{2}{School of Natural Sciences, Institute for Advanced Study, 1 Einstein Drive, Princeton, New Jersey 08540, USA}
\altaffiltext{3}{Department of Astronomy and Astrophysics, University of California, Santa Cruz, CA 95064, USA}
\altaffiltext{*}{Einstein Fellow}

\begin{abstract} 

We perform the first systematic study on how dynamical stellar tides and general relativistic (GR) effects affect
the dynamics and outcomes of binary-single interactions. For this, we have constructed  an N-body code that includes tides
in the affine approximation, where stars are modeled as self-similar ellipsoidal polytropes, and GR corrections using the commonly-used post-Newtonian formalism.
Using this numerical formalism, we are able resolve the leading effect from tides and GR across several orders of magnitude in both stellar radius and
initial target binary separation.  We find that the main effect from tides is the formation
of two-body tidal captures that form during the chaotic and resonant evolution of the triple system. The two stars undergoing the capture
spiral in and merge. The inclusion of tides can thus lead to an increase on the  stellar coalescence rate. We also develop
an analytical framework for calculating the cross section of tidal inspirals between any pair of objects with similar mass. From our analytical
and numerical estimates we find that the rate of tidal inspirals relative to collisions increases as the initial semi-major
axis of the target binary increases and the radius of the interacting tidal objects decreases. The largest effect is therefore found for
triple systems hosting white dwarfs and neutron stars. In this case, we find the rate of highly eccentric white dwarf - neutron star mergers to likely be
dominated by tidal inspirals.  While tidal inspirals occur rarely, we note that they
can give rise to a plethora of thermonuclear  transients such as Ca-rich transients.
\end{abstract}

\section{Introduction}
Stars in dense stellar systems evolve very differently than those  in the solar neighborhood. In an environment of extreme stellar density, like a globular cluster (GC), close encounters between stars are frequent \citep{Heggie:1975uy, Hut:1983js, Hut:1983by, 1992PASP..104..981H, Hut:1993gs, Heggie:1993hi}. 
 Binary-single dynamical encounters, which occur when a single star passes close to a binary and perturbs it, are particularly likely. 
These encounters are much more frequent  than single-single stellar encounters because the orbit of the binary acts as a net -- sweeping passing single stars into periods of resonant triple-object interactivity which are characterized by the formation and dissolution of temporary binary pairings. 
As such, binary-single interactions are responsible for shaping the populations of
close binaries in dense clusters \citep{Heggie:1975uy, 1991ASPC...13..324M, 1992PASP..104..981H, Baumgardt:2002eb, 2005ASPC..328..231I, 2005MNRAS.358..572I, 2009ApJ...707.1533F, 2014A&A...561A..11V}.

Because the energy and momentum are randomized during these chaotic triple interactions, very close passages between pairs of objects are possible \citep[e.g.,][]{1985ApJ...298..502H, 1986ApJ...306..552M, 2006tbp..book.....V}.  During close passages deviations in the dynamics from the behavior of point masses in the
limit of Newtonian gravity can become apparent.  \cite{2014ApJ...784...71S} studied the modification of binary-single dynamics by the inclusion of gravitational wave (GW) energy losses. This work showed a counterintuitive result: passages close enough to modify the dynamics are actually most common in systems involving {\em wide} target binaries. 
We typically associate general relativistic corrections with being most important in compact systems, but \cite{2014ApJ...784...71S} show that the likelihood of generating a very close encounter actually rises as the target binary widens because its larger cross section sweeps in the most perturbing stars. 

This paper focuses on stellar tides, another commonly neglected  physical ingredient  in the full $N$-body equation of motion of stars. Stellar tides can be excited in close passages and much like gravitational radiation, the amplitude of tidally-excited oscillations is very sensitive to the passage distance between two objects.  
Tides play a role in stellar dynamics when stars are sufficiently close that there is an appreciable difference in gravitational force across the object radius.
The relative tidal force scales to leading order as $(R/r)^3$ for a perturber at distance $r$ from a star with radius $R$. The associated
energy transfer has a much steeper dependence on $r$ \citep[][which we refer as PT for the rest of the paper]{1977ApJ...213..183P}.
Thus tides can become important when objects come within a few stellar radii of each other. 

Previous work on how dynamical tides might affect GC evolution has mostly been related to the role of 
binaries that are formed through two-body tidal captures \citep{1975MNRAS.172P..15F, 1977ApJ...213..183P, 1986ApJ...310..176L, 1987ApJ...318..261M}.
These "two-body" binaries have been suggested to, for example, help reverse the
contraction of GC core collapse long before binaries formed by pure
"three-body" interactions are created \citep{1983Natur.305..506K, 1985IAUS..113..347O, 1986ApJ...306..552M}. 
Tidally formed binaries also have potentially observable consequences and were initially suggested
to explain GC X-ray sources \citep{1975MNRAS.172P..15F, 1975ApJ...199L.143C}. Numerical modeling and observations have shown
that few-body interactions must play a role not only in the 
formation and disruption of X-ray binaries \citep{1983Natur.301..587H, 2003ApJ...591L.131P, Pooley:2006ef, 2010ApJ...717..948I, Ivanova:2013tw, 2014A&A...561A..11V}, but of all compact binaries in dense stellar systems \citep[e.g.,][]{Sigurdsson:1993jz, Sigurdsson:1995gh, 2006MNRAS.372.1043I, Ivanova:2008jx}.
This further includes compact neutron star binaries which are believed to be the
progenitors of short gamma-ray bursts  \citep[SGRBs;][]{Grindlay:2006ef,Lee:2010ina}. Other distinct features of dynamically formed binaries
include high eccentricity at merger, which can give rise to a rich variety of electromagnetic and GW
signatures for black holes and neutron stars \citep{2011ApJ...737L...5S, East:2012hg, Gold:2012jg, East:2012es, East:2013iy, 2014ApJ...784...71S}.

The role of dynamical encounters has been further linked to the observed distribution of
pulsars  \citep{2005ASPC..328..147C, 2013IAUS..291..243F, 2014A&A...561A..11V}. In particular, the observed population of
single millisecond-pulsars (MSPs) in GCs suggests that few-body interactions
must happen frequently with outcomes that both assemble and disrupt compact binaries \citep{2013IAUS..291..243F, 2014A&A...561A..11V}. Few-body interactions involving two or more stars are also likely to result in stellar mergers \citep{Fregeau:2004fj}. The remnants
of such mergers have been proposed to partially explain the observed population of so-called blue stragglers \citep[BSs;][]{1953AJ.....58...61S}.
However, the leading formation mechanism of BSs is still
under debate \citep[e.g.,][]{2011MNRAS.416.1410L, 2013MNRAS.428..897L, 2015ebss.book..295K}, with formation mechanisms
ranging from isolated mass transfer \citep{2011Natur.478..356G} to secular dynamics \citep{2009ApJ...697.1048P} and resonant
few-body interactions \citep{Fregeau:2004fj}. 

Despite the clear importance of few-body interactions in shaping the distributions of binaries and singles, no systematic
study has been done of how tidal effects might affect the outcomes. The nature of tidal encounters also remains uncertain. In fact, simulations and semi-analytical models indicate that a tidally formed binary
may lead to a merger rather than
a stable binary \citep{1992ApJ...385..604K, 1993PASP..105..973R}, a concern that was also raised in the
original paper by \cite{1975MNRAS.172P..15F}. The evolution of the tidal capture itself has been studied
using different analytical prescriptions:
\cite{Mardling:1995hx, Mardling:1995it} showed using a linear mode analysis
that if one takes into account the evolving oscillatory state of the stars
on the orbital evolution, tidal captures are likely to undergo quasiperiodic or even random walk behavior.
Similar behavior has also been discussed and seen in work related to the non-linear affine
model developed by \citet{1985MNRAS.212...23C} and  \citet{Luminet:1986cha}, further studied by
e.g. \cite{1992ApJ...385..604K}, \citet{1993ApJS...88..205L,1994ApJ...423..344L,1994ApJ...420..811L, 1994ApJ...437..742L} and \citet{1995ApJ...443..705L},
and later generalized by \citet{2001ApJ...549..467I} and  \citet{2003MNRAS.338..147I}. Non-linear mode couplings could therefore play a role in the problem
of a tidal encounter since the mode excitation spectrum of the star determines the dynamical evolution at subsequent
passages. The outcome is therefore  intimately linked  to the problem of how the energy is dissipated
during the evolution, which remains an open question \citep[e.g.,][]{2014ARA&A..52..171O}. 

GC simulations including approximations of tidal effects have been performed \citep{Mardling:2001dl}, but to our knowledge,
there is little systematic study of how dynamical tides play a role in few-body systems.
Earlier works, particularly by \citet{1985ApJ...298..502H} and \citet{1986ApJ...306..552M},
have discussed the effect of tides and finite sizes. However, their results are based on point-particle simulations and the few-body systems they study
are not evolved consistently with tides. Recent studies by, e.g., \cite{2010MNRAS.402..105G} have modeled a few binary-single interactions
using smoothed particle hydrodynamics (SPH). However, their limited sample mainly consists of very hard target
binaries due to the heavy computational cost of SPH simulations,
and are therefore not representative for the wide distribution of
binary-single interactions that are known to occur in dense stellar systems.

The first clear insight  about how tides might affect binary-single
interactions in general, was given by \citet{1992ApJ...385..604K}, who correctly suggested
that tidal effects in three-body interactions will lead to two-body tidal captures during the chaotic evolution of the
triple system. Furthermore, \citet{1992ApJ...385..604K} imagined that these tidally formed binaries probably have an orbital distribution different from those formed
by two-body captures in the field. 

In our work we study close three-body encounters using an $N$-body prescription where the orbital dynamics
is evolved consistently with both tides and GR. We show that the main effect of including tides is the formation of two-body tidal captures during the
three-body evolution, in agreement with earlier predictions by \citet{1992ApJ...385..604K}. Using both numerical and analytical arguments we illustrate
that the relative rate of these tidal captures increases as the radius of the tidal object decreases relative to the size of the initial target
binary. In the astrophysical  context, tides therefore show the largest effect when the perturbed  object is a white dwarf.

Our analysis of the role of tides in binary-single dynamics proceeds as follows.
We discuss the general properties of binary-single encounters and build some intuition for the possible role
of tides in shaping these encounters in Section \ref{sec:Binary-Single Interactions and Tidal effects}.
In Section \ref{sec: N-body with Tides and GR: Numerical Methods} we describe a numerical formalism for including tidal excitation in $N$-body encounters
by treating stars as compressible ellipsoids. In Section \ref{sec: Numerical Scattering Results} we describe results of scattering experiments of
large numbers of binary-single encounters. In Section \ref{sec:Analytical Models} we derive analytical relationships to interpret the dependence of these results on stellar type and binary properties. Finally, in Section \ref{sec:Discussion} we discussed  our findings while our  conclusions are presented in Section \ref{sec:Conclusion}.

\section{Binary-Single Interactions and Tidal effects}\label{sec:Binary-Single Interactions and Tidal effects}

We begin our exploration of tidal effects in binary-single stellar interactions by reviewing some of the basic properties of these encounters in the point-mass limit approximation. We then consider the possible role of finite-size  objects in shaping the nature of the interactions  and the range of possible outcomes.

\subsection{Binary-Single Interactions}\label{sec: Basics of Binary-Single Interactions}

A binary in a dense stellar environment is subject to weak and strong perturbations from the surrounding stars.
In a typical dense stellar system, many of these perturbing stars will be single stars \citep{Heggie:1975uy, 1992PASP..104..981H}.
The resulting binary-single stellar encounters happen at a rate which depends
on the semimajor axis (SMA) of the binary, $a_{0}$, the total mass of the three interacting objects, $m_{\rm tot}$, and the relative velocity of the perturbers at infinity, $v_\infty$
(the velocity dispersion of the stellar system).
A larger SMA makes the binary a larger target for interloping objects, and a larger total mass enhances gravitational focusing towards the binary.  

In this paper, we focus on close binary-single encounters: those in which the single object reaches a pericenter distance -- with respect to the center of mass (COM) of the binary --
of approximately the binary SMA $a_{0}$ \citep{2014ApJ...784...71S}. 
The cross section has contributions  from the  geometrical scale of the binary, $\approx \pi a_{0}^2$, and from gravitational focusing. In the limit when gravitational
focusing dominates, the close interaction (CI) cross section can be written as \citep{{2014ApJ...784...71S}},
\beq
\sigma_{\rm CI} \approx \frac{2 \pi G m_{\rm tot} a_{0} }{v_\infty^2}.
\label{eq:sigma_CI}
\eeq
The corresponding event rate of close interactions experienced by this binary is given by
\beq
\Gamma_{\rm CI} \approx n \sigma_{\rm CI} v_\infty,
\eeq
where $n$ is the number density of perturbing stars.

Gravitational focusing is dominant in establishing the binary cross section when the binary is {\em hard} relative to the surrounding stellar system.  Hard binaries are those that are sufficiently compact that the net energy of a typical three body encounter is negative \citep{Hut:1983js}.
This is fulfilled when $v_{\infty}$ is less than the characteristic velocity of the binary, $v_{\rm c}$, defined as \citep{Hut:1983js}
\begin{equation}
v_{\rm c}^2 \equiv G\frac{m_1m_2(m_1+m_2+m_3)}{m_3(m_1+m_2)}\frac{1}{a_{0}},
\label{eq:v_c}
\end{equation}
where 1 and 2 refer to the two stars initially in the binary, and 3 to the incoming perturber.
In the opposite case, {\em soft} binaries have net positive energy when the energy of a typical perturber is included. 
\cite{Heggie:1975uy} has shown that (as a result of their energetics) hard binaries tend to persist and tighten in dense stellar
systems, while soft binaries tend to dissolve. 
The maximum SMA a binary can have without dissolving, denoted $a_{\rm HB},$ is therefore set by the limit where $v_{\rm c} \approx v_{\infty}$.
In the equal mass case, we see from Equation \eqref{eq:v_c} that $a_{\rm HB} \propto m/v_{\infty}^2$.

Because the net energy is negative in an encounter involving a hard binary (HB), the triple system formed by the binary and the perturbing single object may pass through many iterations before  a final configuration is attained. We call these multi-passage encounters {\em resonant} interactions.  They are characterized by the formation and disruption of many intermediate state (IMS) binaries before an eventual outcome \citep{2014ApJ...784...71S}. In these chaotic interactions, memory of the initial configuration is lost and all objects are therefore equally likely to be ejected (if they are identical).

In general, several possible outcomes result from a binary-single close interaction. In the limit of Newtonian point masses, these include:
\begin{itemize}
\item {\em ionization} of the system (all three objects are mutually unbound). This outcome can occur  in the soft-binary regime  but not  in the HB limit. 
\item a {\em fly-by} encounter, in which the binary survives the encounter but its original  orbit  is modified as a result.
\item an {\em exchange} encounter, in which the perturber exchanges into the  binary system, ejecting one of the original binary components. 
\end{itemize}

In GR, an additional outcome is possible:
\begin{itemize}
\item an {\em inspiral} between two compact objects through the emission of
GW radiation \citep{2014ApJ...784...71S}.
\end{itemize}

These inspirals occur when an IMS binary is generated that has very high eccentricity (and thus small pericenter distance) so that GW losses strongly modify the equation of motion. \cite{2014ApJ...784...71S} found that this configuration arises primarily in resonant interactions involving hard binaries, where many opportunities for close passages occur before the binary-single encounter concludes.

\begin{figure}
\centering
\includegraphics[width=\columnwidth]{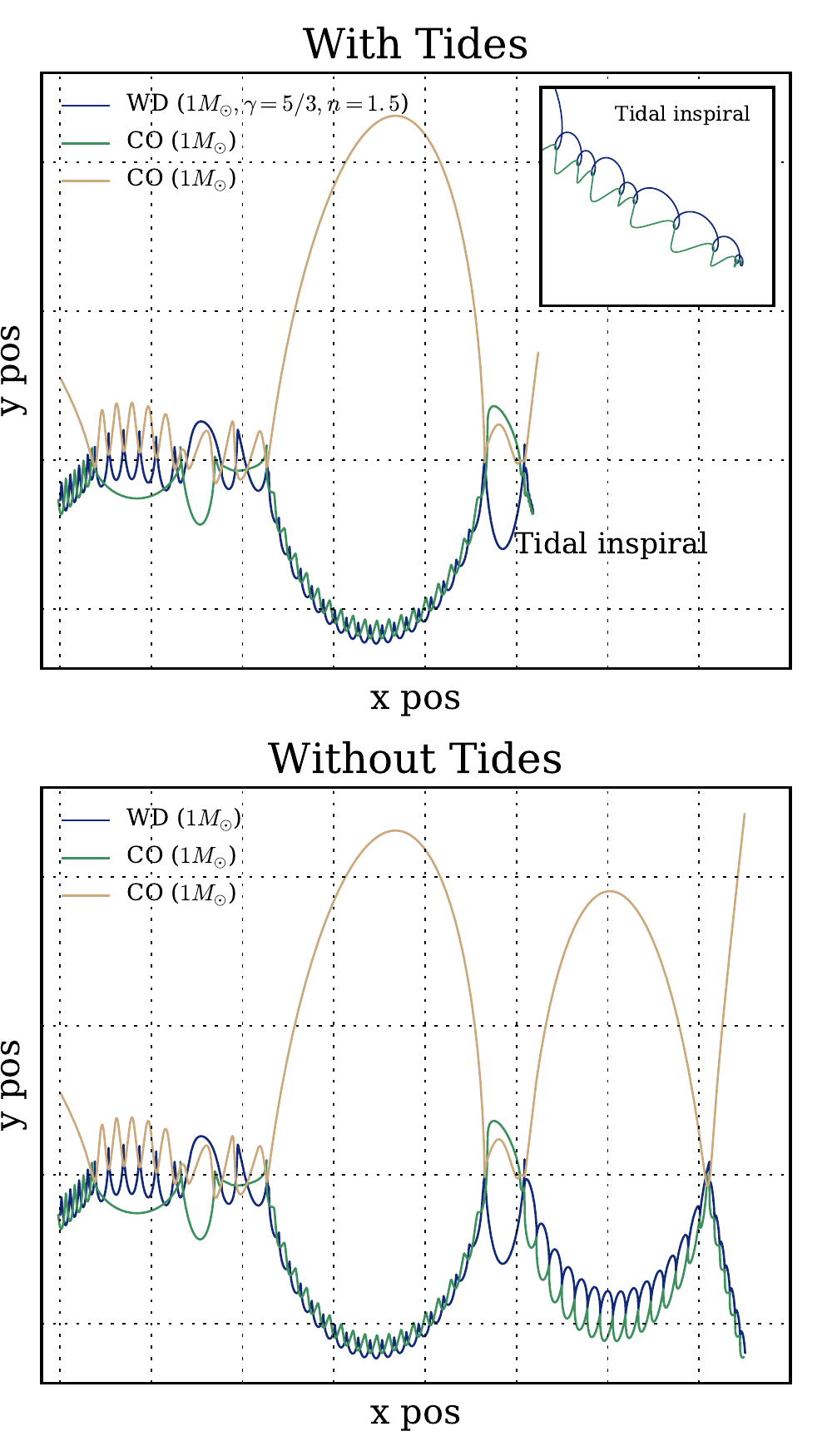}
\caption{
Orbital evolution of a binary composed of a white dwarf (WD, blue) and a compact object (CO, green), interacting with a single incoming
compact object (CO, brown). The interactions propagate from left to right.
The \emph{top plot} shows the evolution if tidal effects are included in the equation of motion, where
the \emph{bottom plot} shows the evolution if the WD is modeled as a simple solid sphere. Both simulations
have the exact same initial conditions. Each object has a mass of $1M_{\odot}$,
the radius of the WD is set to $0.0086R_{\odot}$ and is  modeled as a polytrope with adiabatic index $\gamma = 5/3$
and polytropic index $n=1.5$ (see Section \ref{sec: N-body with Tides and GR: Numerical Methods}). The initial semimajor axis is $5\cdot10^{-3}$ AU.
As seen in the top plot, when tidal effects are included a close passage between two of the three stars can lead to
a tidal capture during the chaotic evolution of the system (indicated by the label {\it Tidal inspiral}).
In this paper we denote such a capture a \emph{tidal inspiral}
mainly due its similarity to GW inspirals \citep{2014ApJ...784...71S}.
The insert box at the upper right shows a zoom in on the final few orbits of the inspiral between the WD and the CO.
As seen, the orbital evolution is not smooth which is due to the strong coupling between the tidal modes and the orbit.
The final stage of the inspiral is a coalescence between the WD and the CO.
}
\label{fig:com3body_WT_NT}
\end{figure}

\subsection{Finite Size Objects in Binary-Single Interactions: Collisions and Tides}

Having outlined the basic properties of binary-single encounters in dense stellar systems, we move on to consider the structure of the  stellar  objects involved. 
If objects have finite size, rather than being point masses, the most obvious new outcome is that two objects may undergo a \emph{collision} if they approach each other with pericenter distance less than the sum of their radii \citep{1985ApJ...298..502H}. 

This paper focuses primarily on a second effect: close passages may also excite non-radial oscillations in stars originating from the difference in gravitational force between the stellar COM and the stellar limb. 
Tidal forces distort stars during particularly close passages (those with pericenter distance within a few stellar radii for equal mass objects). 
When a distorted star flies free of this perturbing force, it begins to oscillate (non-radially) around its equilibrium configuration.  The  quadrapolar fundamental mode is the primary oscillation mode excited with $l=2$, $m=\pm2$ in spherical
harmonic notation \citep{1977ApJ...213..183P, 1986ApJ...310..176L, 1992ApJ...385..604K}.  
The energy and momentum carried by these oscillations comes at the expense of the orbital motion of the stars. Thus, exciting tidal oscillations drains energy and  angular momentum from the orbit of a pair of objects and -- by depositing that energy and momentum elsewhere -- tidal oscillations can act as a sink term. 

By modifying the equation of motion, the inclusion of tides in binary-single encounters has many potential effects on the dynamics and distribution of outcomes. 
One particularly dramatic outcome is the possibility for a  new interaction channel. 
In some encounters, IMS binaries form which could spiral in from an initially wide orbit toward merger, by transferring orbital energy into tidal oscillations.
These tidal capture events, or {\em tidal inspirals}, as we will refer to them, occur in cases where an IMS binary has high eccentricity and undergoes a very close pericenter passage, leading to strong tidal forcing. 

Figure \ref{fig:com3body_WT_NT} shows two binary-single interactions, each with the same initial conditions. The simulation shown on the {\it top panel}  includes tides, while the simulation shown on the {\it bottom panel} does not (we use a numerical methodology in integrating these encounters which will be discussed in Section \ref{sec: N-body with Tides and GR: Numerical Methods}). 
In this particular example we clearly see that a tidal inspiral has formed when
tides are included. 
The formation of an eccentric IMS binary leads to a close, strong  tidal interaction. In this way, tidal inspirals are similar to GW inspirals, which also form in rare, very close pericenter passages during resonant interactions. This similarity will be exploited in Section \ref{sec:Analytical Models}, in order to derive a simple yet generalized analytical understanding of such encounters.

Our study focuses on understanding the formation rate of these tidal inspirals, and classifying their effects on the triple dynamics of binary-single encounters. 
In the subsection below we present some of the characteristic scalings
involved in the problem, which can lead  to a simple understanding of what combinations of
orbital and object parameters can successfully generate tidal inspirals.

\subsection{Tidal Captures from Simple Scaling Relations}\label{sec:Tidal Captures from Simple Scaling Relations}

Here we derive a few fundamental scaling relations in order to illustrate and build intuition
for when tides are expected to play a leading role in binary-single interactions.
A naive guess is that tides are expected to play a role when the SMA of the target
binary, $a_{0}$, is relatively small compared to the radius of the stars, $R$.
However, in the following we show that the dependence is
indeed the opposite: tidal outcomes are maximized when $a_{0}$ is large compared
to $R$. 

For tidal perturbations to alter the evolution of a chaotic binary-single interaction, the energy
deposited into tidal oscillations during a close passage between two of the
three stars, $\Delta{E_{\rm tid}}$, must be similar to the total orbital energy of the  system, $E_{0}$.
Assuming the tidal energy transfer falls off as a simple power-law with pericenter distance, $r_{\rm p}$, and that the
orbital energy is dominated by the target binary (HB limit), the energy terms can be written as
\begin{equation}
\Delta{E_{\rm tid}} \propto \frac{m^2}{R} \left(\frac{R}{r_{\rm p}}\right)^{\beta} {\rm and} \ \  {E_{0}} \propto \frac{m^2}{a_{0}},
\end{equation}
where $\beta \gtrsim 6$ as we'll discuss later. By equating $\Delta{E_{\rm tid}}$ and $E_{0}$, we can now solve for the
pericenter distance at which two stars must pass for tides to have a substantial
energetic effect on the triple system. We call this distance
$r_{\rm tid}$, and it scales as 
\begin{equation}
r_{\rm tid} \propto R\left(\frac{a_{0}}{R}\right)^{1/\beta} \propto R\left(\frac{E_{\rm star}}{E_{0}}\right)^{1/\beta},
\label{eq:r_tid_simple_scaling}
\end{equation}
where $E_{\rm star}\propto m^{2}/R$ is the binding energy of the star. The tidal pericenter distance $r_{\rm tid}$ is not merely
a constant times $R$, as has been assumed in many previous studies \citep[e.g.][]{Fregeau:2004fj}:
$r_{\rm tid}$ depends on the total energy of the system, $E_{0}$. In particular, $r_{\rm tid}$ increases
with $a_{0}$ in contrast to $R$. As a result, the fraction of tidal encounters ($r_{\rm p} < r_{\rm tid}$)
also increases relative to the fraction of collisions ($r_{\rm p} < R$). This leaves open the possibility of tidal encounters dominating over 
collisions. Finally, we note that our scalings for $r_{\rm tid}$ are similar to the well known scalings for the single-single tidal capture
radius \citep{1975MNRAS.172P..15F}, but with $E_{0}$ replaced by $E_{\infty} \propto mv_{\infty}^2$, where $v_{\infty}$ is the velocity dispersion of the stellar system.

Motivated by the scalings in Equation \eqref{eq:r_tid_simple_scaling}, we explore the effect of tides as a function of orbital properties and object compactness.
We do this by first using full numerical simulations which include both tidal and GR correction terms. We describe our
approach and results in Sections \ref{sec: N-body with Tides and GR: Numerical Methods} and \ref{sec: Numerical Scattering Results}, respectively. 
This is then followed by an analytical derivation in Section \ref{sec:Analytical Models} aimed at understanding how tidal and GW captures are formed
as a function of target binary separation and object compactness.

\begin{figure}
\centering
\includegraphics[width=\columnwidth]{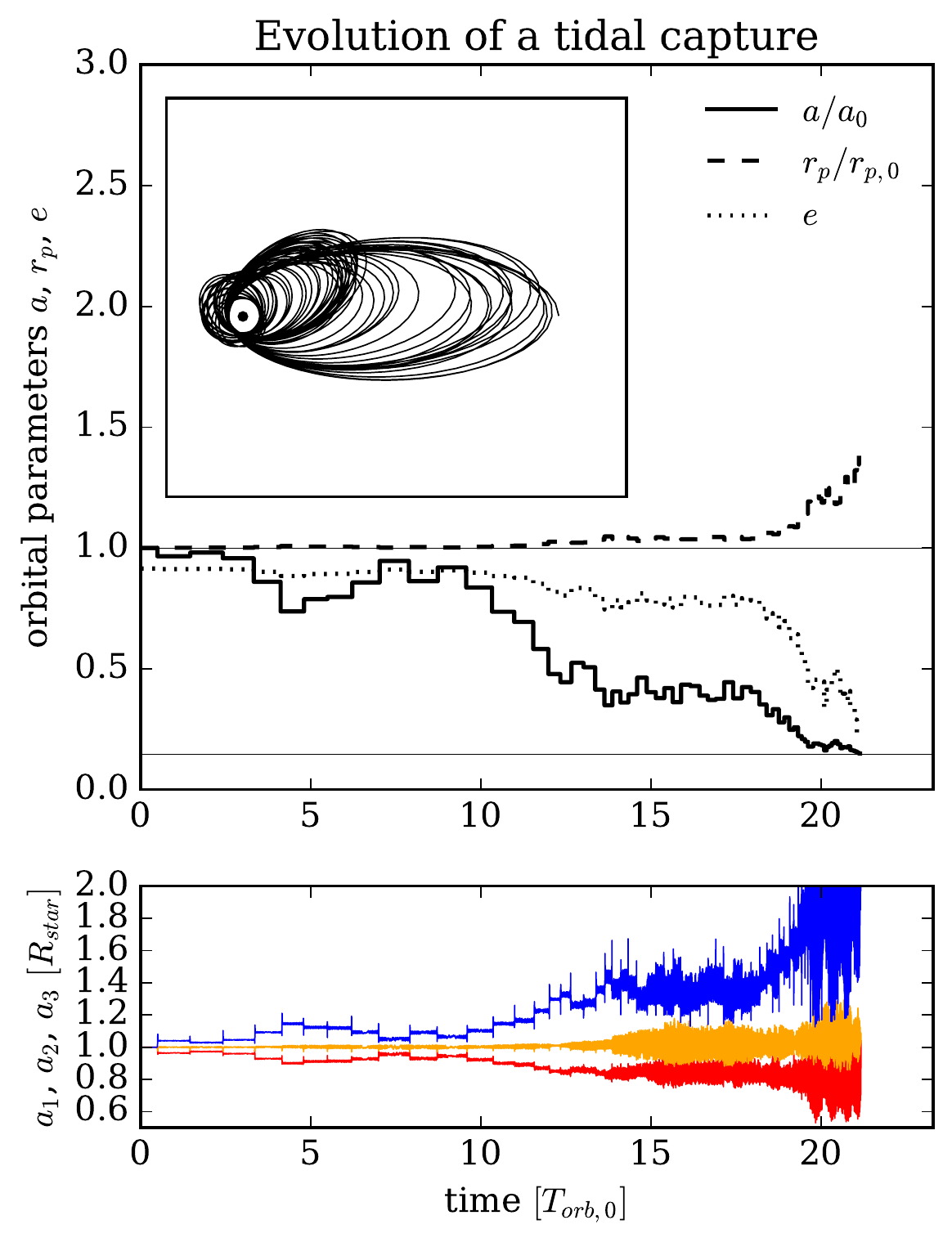}
\caption{Evolution of a strong tidal capture with initial pericenter $r_{\rm p} \sim 2R_{\odot}$ and semimajor axis $0.1$AU, between a
neutron star ($1.4M_{\odot}$) and a solar type star ($1M_{\odot},\ 1R_{\odot},\ \gamma=5/3,\ n=3$). The simulation is performed
with our new $N$-body code which includes tidal effects using the affine model \citep{1985MNRAS.212...23C}. This model allows for non-linear
ellipsoidal stellar deformations, which are consistently evolved and coupled to the orbital evolution of the $N$-body system.
\emph{Top}: Evolution of semimajor axis $a$, pericenter $r_{\rm p}$, and corresponding eccentricity $e$,
as a function of time for the evolving binary. The change in the parameters results from the orbit sinking
energy and angular momentum into the tidal modes.
As can be seen, the orbital
parameters are performing a quasi-random walk due to the short time scale of the tidal oscillations  relative to the
binary's orbital time. The \emph{upper left} panel shows the trajectory of the neutron star relative to the star.
\emph{Bottom}: Corresponding time evolution of the three principal axis of the ellipsoidal model of the star. The star is quite distorted at the end of the encounter
and the outcome is therefore likely to be a partial disruption.
}
\label{fig:twobody_tidalcapture_NSMS}
\end{figure}

\section{Numerical Methods: $N$-body with Tides and General Relativistic Corrections}\label{sec: N-body with Tides and GR: Numerical Methods}

For this study we constructed  a new $N$-body code which includes GR effects
and dynamical tides. The GR corrections are modeled using the \emph{post-Newtonian} (PN) expansion formalism \citep{Blanchet:2006kp},
while tides are dynamically evolved using the \emph{affine model} of compressible ellipsoidal stars, which allows
for the implementation of non-linear stellar deformations  \citep{1985MNRAS.212...23C}. To evolve this system we make use of the \texttt{ODEPACK}, \texttt{LAPACK} and \texttt{GSL} libraries.
$N$-body codes including tides at the linear PT level are available \citep{Mardling:2001dl}, however, it is essential to our study
to have a fully dynamical model since our interactions often undergo several strong, and sometimes simultaneous, encounters
between two or more objects. The affine model also performs better than the PT model during very close passages, which is important since these
are associated with the largest transfer of energy and angular momentum between the tides and the binary's orbit.

We numerically estimate outcome cross sections using a standard Monte Carlo (MC)
approach \citep{Hut:1983js}, which in our case, makes use of an  MPI parallelized version.
In the following sections we describe in further detail our GR and tidal implementations
in the $N$-body code. Regarding notation, we use boldface below for denoting vectors and matrices.

\subsection{General Relativistic Corrections}

We model the effects from GR, such as GW radiation, using the PN formalism \citep{Blanchet:2006kp}. In this formalism GR is included by simply adding
correction terms of order $v/c$ to the classical newtonian acceleration $\propto 1/r^{2}$. Following this framework, the acceleration of an object with
mass $m_{1}$ due to an object with mass $m_{2}$ can be written as

\begin{equation}
\acc=\acc_{0} + c^{-2}\acc_{2} + c^{-4}\acc_{4}+c^{-5}\acc_{5}+\mathcal{O}(c^{-6}),
\label{eq:pn_overview}
\end{equation}
where the standard Newtonian acceleration, $\acc_{0}$, is given by
\begin{equation}
\acc_{0} = - \frac{G m_2}{r_{12}^2} {\bf \hat r}_{12}.
\end{equation}
Here we have defined the separation vector ${\bf r}_{12}  = {\bf r}_1 - {\bf r}_2$, where its magnitude is $r_{12} = |{\bf r}_{12}|$ and its direction is ${\bf \hat r}_{12} = {\bf r}_{12}/r_{12}$.
The terms, $\acc_{2}$ and $\acc_{4}$, conserve energy and account for the periastron shift. These terms play a key role when
describing hierarchical secularly evolving systems \citep[e.g.][]{2013ApJ...773..187N}.
The leading order term that represents the energy and momentum loss carried away by GW radiation is $\acc_{5}$, often denoted the 2.5PN term. 
For our study, the 2.5PN correction is the only relevant term to include since any effects from the lower order precession terms will simply average out when calculating the cross section from a large number of scatterings \citep{2006ApJ...640..156G, 2014ApJ...784...71S}.
The 2.5PN term takes the following form
\begin{equation}
\begin{split}
\acc_{5} =	&  \frac{4}{5}\frac{G^{2}m_{1}m_{2}}{r_{12}^{3}}    \left[  \left ( \frac{2 Gm_1}{r_{12}}  - \frac{8 Gm_2}{r_{12}} - v_{12}^2 \right )  {\bf v}_{12} \right. \\
  		& \left. + \ ( {\bf \hat r}_{12} \cdot  {\bf v}_{12}  )\left ( \frac{52 Gm_2}{3 r_{12}} - \frac{6 Gm_1}{r_{12}} + 3v_{12}^2   \right )   {\bf \hat r}_{12}  \right],
\end{split}
\label{eq:25PNterm}
\end{equation}
where the relative velocity scalar, $v_{12}$, and vector, ${\bf v}_{12}$, are defined following the same conventions as in \citet{Blanchet:2006kp}. 
Further details can be found in \citet{2014ApJ...784...71S}, where we have  shown that the main effect from this correction is the formation of binaries that
inspiral due to GW emission during the resonant interaction.

\subsection{Tidal Corrections}
We dynamically evolve stellar tides using the affine model developed in \citet{1985MNRAS.212...23C}. In this model, the tidal response and the coupling to the orbit
is calculated assuming the star can be described by a triaxial object composed of self-similar ellipsoids.
The model accounts for different radial density and pressure profiles for the distorted star. As a result, one can self-consistently evolve close encounters between stars
with different polytropic index, mass and radius.
The advantage of the model is that it accounts for  the non-linear deformations of the star along its triaxial axes, while its disadvantage is that it only
accounts for the evolution of the $l=2$ mode. However, as pointed out by, for example,  \citet{1992ApJ...385..604K}, the $l=2$ $f$-mode is expected to dominate the energy transfer.
This makes the affine model a great choice for evolving highly dynamically $N$-body systems. Work by \citet{1994ApJ...423..344L} has also compared its validity  to full
SPH simulations. More complicated versions of the model, which incorporate mass loss and allow for the principle axis to
evolve independently throughout the star  have been constructed \citep{2001ApJ...549..467I, 2003MNRAS.338..147I}. Yet,
the corrections in these models  come at a very high computational cost.

In what follows we briefly discuss the basic
evolution equations of the affine model, but  refer the reader to the extensive literature on the subject  for more details \citep{1985MNRAS.212...23C, Luminet:1986cha, 1992ApJ...385..604K, 1992MNRAS.258..715K, 1993ApJS...88..205L, 1993ApJ...406L..63L, 1994ApJ...420..811L, 1994ApJ...423..344L, 1995ApJ...443..705L, 1995MNRAS.275..498D, 1996PThPh..96..901O, 2001ApJ...549..467I, 2003MNRAS.338..147I}.
In order to validate our code formalism, we compared our results against the 2-body encounter outcomes  presented  in \citet{1994ApJ...437..742L}.

In the affine model the position of a fluid element inside the star at time $t$ is given by
\begin{equation}
x_{i}(t) = q_{ij}(t)\hat{x}_{j},
\end{equation}
where $q_{ij}$ is a 3x3 matrix, $\hat{x}_{j}$ is the fluid position at time $t=0$, and summation over repeated indices is assumed.
The tidal evolution of the star is completely determined
by the evolution of the components of $q_{ij}$. The evolution of a star experiencing  a deformation $\bf{q}$ and its  respective coupling to other stars and their mutual orbital
motion can found using the Lagrangian formalism. In our case, the Lagrangian of the full $N$-body system is given by
\begin{equation}
L = L_{\rm I} + L_{\rm E},
\end{equation}
where $L_{\rm I}$ is the \emph{internal} Lagrangian which relates to the internal energy of the individual stars, and $L_{\rm E}$ is the \emph{external} Lagrangian
which is composed of the kinetic COM energy and the potential energy between the interacting stars.
The internal Lagrangian is given by 
\begin{equation}
L_{\rm I} = T - U - \Omega
\end{equation}
where $T$ is the stellar kinetic energy, $U$ is the gas energy, and $\Omega$ is the self-gravitational potential energy. Each of these energy terms
take the following form for a single star in the affine model,
\begin{equation}
T = \text{Tr}[T_{ij}],\ T_{ij}= \frac{1}{2}\dot{q}_{ia}\dot{q}_{ja}\mathcal{M}_{*}
\end{equation}
\begin{equation}
U = - \frac{\Omega_{*}}{3(\gamma-1)}|{\bf{q}}|^{1-\gamma}
\end{equation}
\begin{equation}
\Omega = \text{Tr}[\Omega_{ij}], \ \Omega_{ij} = \frac{1}{2}\Omega_{*}|{\bf{S}}|^{-1/2}{\bf{A}}{\bf{S}}
\end{equation}
where $\text{Tr}[Y_{ij}]$ denotes the trace of matrix $Y_{ij}$, |{\bf{Y}}| denotes the determinant of ${\bf{Y}}$, $\dot{q}$ is the time derivative
of $q$, ${\bf{S}} = {\bf{q}}{\bf{q}}^{\rm T}$ where $\rm T$ is here the transpose, $\mathcal{M}_{*}$ is the scalar quadrupole moment
of the spherical star\footnote{A table with calculated values for stars with
different polytropic index $n$ is given by \cite{1995MNRAS.275..498D} in Table A1.}, $\gamma$ is the adiabatic index
of the stellar fluid, and
\begin{equation}
\Omega_{*} = - \frac{3}{5-n}\frac{Gm^{2}}{R},\;{\rm and}
\end{equation}
\begin{equation}
{\bf{A}} = |{\bf{S}}|^{1/2} \int_{0}^{\infty}du\frac{\left({\bf{S}}+u{\bf{I}}\right)^{-1}}{|{\bf{S}} + u{\bf{I}}|^{1/2}}.
\end{equation}
Here the matrix ${\bf{I}}$ is the identity matrix, $u$ denotes an integration variable,
$m$ and $R$ denotes the mass and the radius of the star, respectively, and $n$ the stellar polytropic index.
The elliptical integral needed for calculating
${\bf{A}}$ has to be performed at each time step\footnote{A significant improvement in speed can be achieved by solving the integral in
coordinates where ${\bf{S}}$ is diagonal, in this case the integral
reduces to the Carlson's incomplete integral of the third kind \citep{1992ApJ...385..604K}.}.
The external Lagrangian is given by 
\begin{equation}
L_{\rm E} = K - \Phi,
\end{equation}
where $K$ is the COM kinetic energy and $\Phi$ is the total potential energy of the system.
The term $\Phi$ is in the case of two stars, $1$ and $2$, found by
integrating the potential of star $1$ over the density distribution of star $2$ and viceversa. To quadrupole order
this term takes the form,
\begin{equation}
\Phi_{12} = - \frac{m_{1}m_{2}}{|X|} - \frac{1}{2}C_{ij}\left[ m_{1}\mathcal{M}_{*}^{(2)}S_{ij}^{(2)} +  m_{2}\mathcal{M}_{*}^{(1)}S_{ij}^{(1)} \right]
\end{equation}
where $C_{ij}$ is the tidal tensor here defined as
\begin{equation}
C_{ij} = \frac{3X_{i}X_{j} - |X|^{2}I_{ij}}{|X|^{5}},
\end{equation}
and ${\bf{X}}$ is the relative position vector between the two stars. The final $L_{\rm E}$ is given by the sum of the individual kinetic terms
and the pairwise potential terms.
From applying the Lagrange formalism, the equation of motion (EOM) of the deformation matrix $\bf{q}$ for a single star is given by
\begin{equation}
\ddot{q}_{ia} = q_{aj}^{-1}\left( \Omega_{ij} - |{\bf{q}}|^{1-\gamma}\Omega_{*}I_{ij}/3 \right)/\mathcal{M}_{*} + \sum_{n}m_{n}C_{ij}^{(n)}q_{ja},
\end{equation}
where the COM acceleration term arising from tidal couplings, here denoted $\acc_{\rm TC}$, is found to be
\begin{equation}
\acc_{\rm TC} = \frac{1}{2m} \sum_{n} Q_{ijk}^{(n)} \left( m\mathcal{M}_{*}^{(n)}S_{ij}^{(n)} +  m_{n}\mathcal{M}_{*}S_{ij} \right).
\end{equation}
The tensor $Q_{ijk}$ is here the derivative of the tidal tensor $C_{ij}$ with respect to the relative COM coordinate $X_{k}$, which we find takes the following form,
\begin{equation}
Q_{ijk} = \frac{3X_{k}}{{|X|^{5}}} \left[ \frac{\partial X_{i}X_{j}}{\partial X_{k}} \frac{1}{X_{k}} - 5 \frac{X_{i}X_{j}}{{|X|^{2}}} + I_{ij} \right].
\end{equation}
For the sums over index $n$ in the equations above, one has to sum over all stars or objects in the $N$-body system
except for the star in question.

We do not include tidal energy and angular momentum dissipation within oscillating stars for the results presented
in this paper --  tidally induced stellar oscillations are as a result not damped, and can therefore exchange energy and angular momentum with
the COM orbital motion throughout the full interaction. This can result in long term chaotic behavior (See Figure \ref{fig:twobody_tidalcapture_NSMS}).

The non-dissipative assumption formally corresponds to the limit where the characteristic orbital time is less than the dissipation time \citep[see Section 3 in][for a discussion on how a tidal capture undergoing multiple passages depends on the relative values for the viscous, orbital and Kelvin-Helmholtz time-scales]{1992MNRAS.255..276N}. 
The main reason for making this assumption, is simply that the non-linear tidal dissipation is still very poorly understood.
It is nonetheless  possible to include an estimate for the dissipation in the affine model by introducing a shear viscosity term \citep{1994ApJ...437..742L}.
However, it is highly uncertain how and where the  heat deposition  occurs within the perturbed star. One consequence of dissipation is expansion
followed by mass loss of the star, which could lead to a tidal runaway. As a result,  one would expect more tidal inspirals
to form in our simulations if dissipation is swift. The results we report here thus serve as strict  lower limit to the tidal inspiral rate and should be considered as such.
Several studies have been done on damping and dissipation \citep{1987ApJ...318..261M, 1992ApJ...385..604K, 1996ApJ...466..946K, 2014ARA&A..52..171O}, and we
are currently working on prescriptions for including both dissipation and mass loss into the affine model\footnote{Mass loss is possible in the extended
model presented in \citet{2001ApJ...549..467I} and \citet{2003MNRAS.338..147I}. For an implementation of linear tides in $N$-body codes we refer the reader to  \citet{Mardling:2001dl}.}.

For our analytical estimation of the tidal inspiral cross section, which is presented in Section \ref{sec:Analytical Models},  we assume that the orbital energy loss
during each pericenter is constant. This corresponds to the limit where the tidal energy is fully radiated away from the system between each
subsequent passages. The overall agreement between this analytical model and our simulations, shown in Figure \ref{fig:cs_all}, indicates 
that dissipation will probably not change the relative scalings of the inspiral cross section and it is  likely that only the normalization will be affected.
We expect that our upcoming  work will shed light on this matter.

\subsubsection{Tidal Capture Example}\label{sec:Tidal Capture Example}

An example of a tidal capture between a neutron star and a solar type star
is shown in Figure \ref{fig:twobody_tidalcapture_NSMS}. The chaotic evolution of the SMA (solid black line) is due to
the tidal mode-orbit coupling that is consistently modeled in the affine model. Since we don't include dissipation, the two stars
are not able to merge and are unable to come closer than their initial pericenter distance $r_{\rm p}$. Angular momentum is stored in the
tidal modes and in the affine model, this scales with the corresponding change in energy $\Delta{E}$ by \citep{1992ApJ...385..604K},
\begin{equation}
\Delta{L} \approx \Delta{E}\sqrt{{15\mathcal{M}_{*}}/{|\Omega_{*}|}}.
\end{equation}
However, this change is very small compared to the initial orbital angular momentum $L_{0}$.
If we for example consider a tidal capture from an orbit with
$a_{0}>r_{\rm p}$, one can easily  see that the change $\Delta{L}$ relative to $L_{0}$ is small and it is given by  $\Delta{L}/L_{0}\propto (R/a_{0})$.
The analytical solution to the stellar separation
at circularization ($e=0$) when including the $\Delta{L}$ correction term is shown by the lower horizontal solid black line in
the upper plot in Figure \ref{fig:twobody_tidalcapture_NSMS} (slightly lower than $2r_{\rm p}$).

We find that for some strong tidal captures the two stars merged once their orbit is circularized. This is
due to the tidal acceleration term  $\acc_{\rm TC}$,  which for small binary separation becomes steeper than the point mass
potential and  causes the instability \citep[similar to the innermost stable orbit in GR,][]{1995ApJ...443..705L}. While we easily
identified when this happens,  understanding the outcome of such strong tidal interactions  requires the use of full hydro simulations. 
For this reason, in our $N$-body experiments we keep track of binaries that circularize (tidal inspirals) and not only binaries
that merge due to this instability.
The dynamical instability of close binaries in the
affine limit was extensively explored by \citet{1992ApJ...385..604K} and \citet{1994ApJ...423..344L}.

\begin{figure}
{\includegraphics[width=\columnwidth]{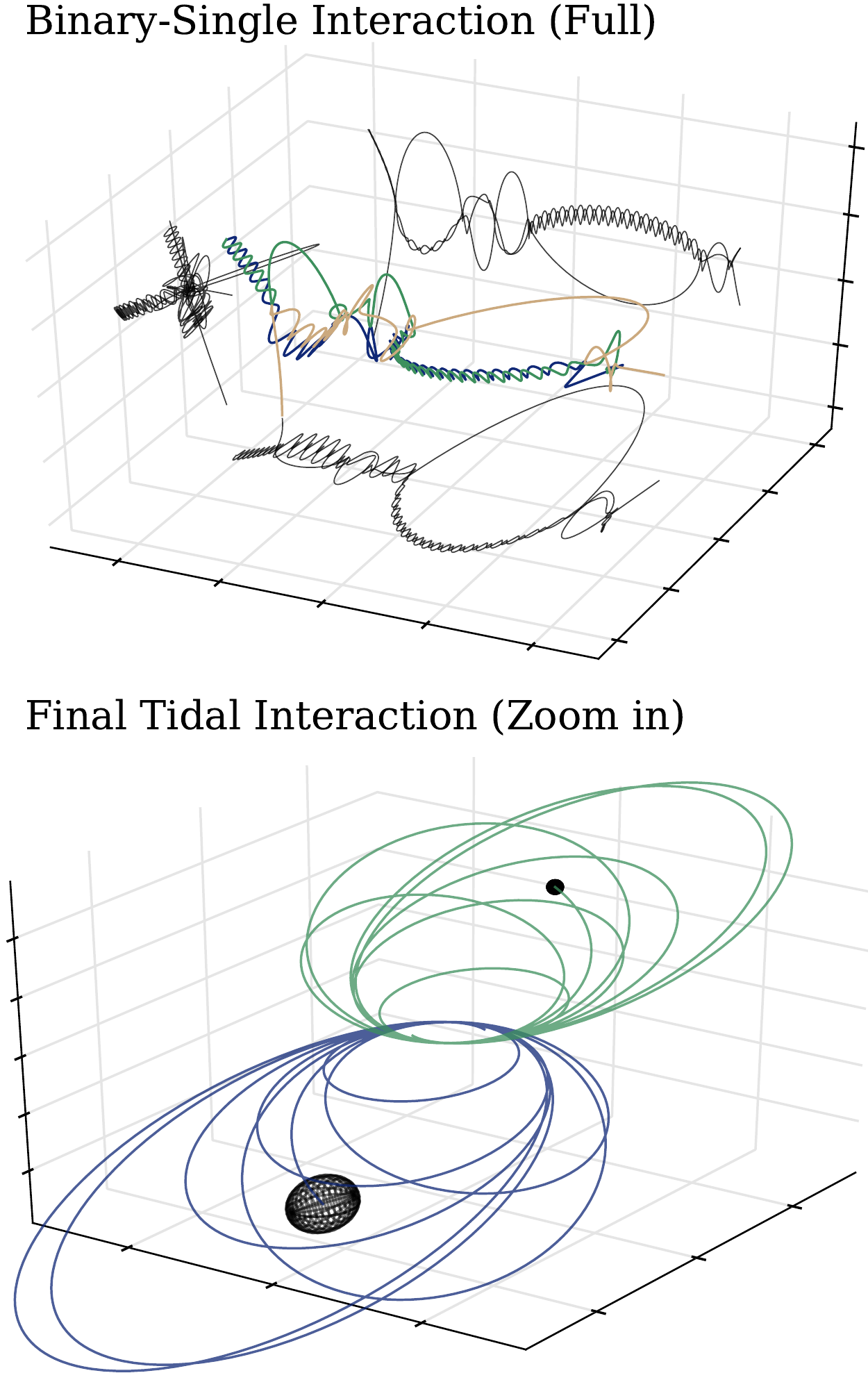}}
{\includegraphics[width=\columnwidth]{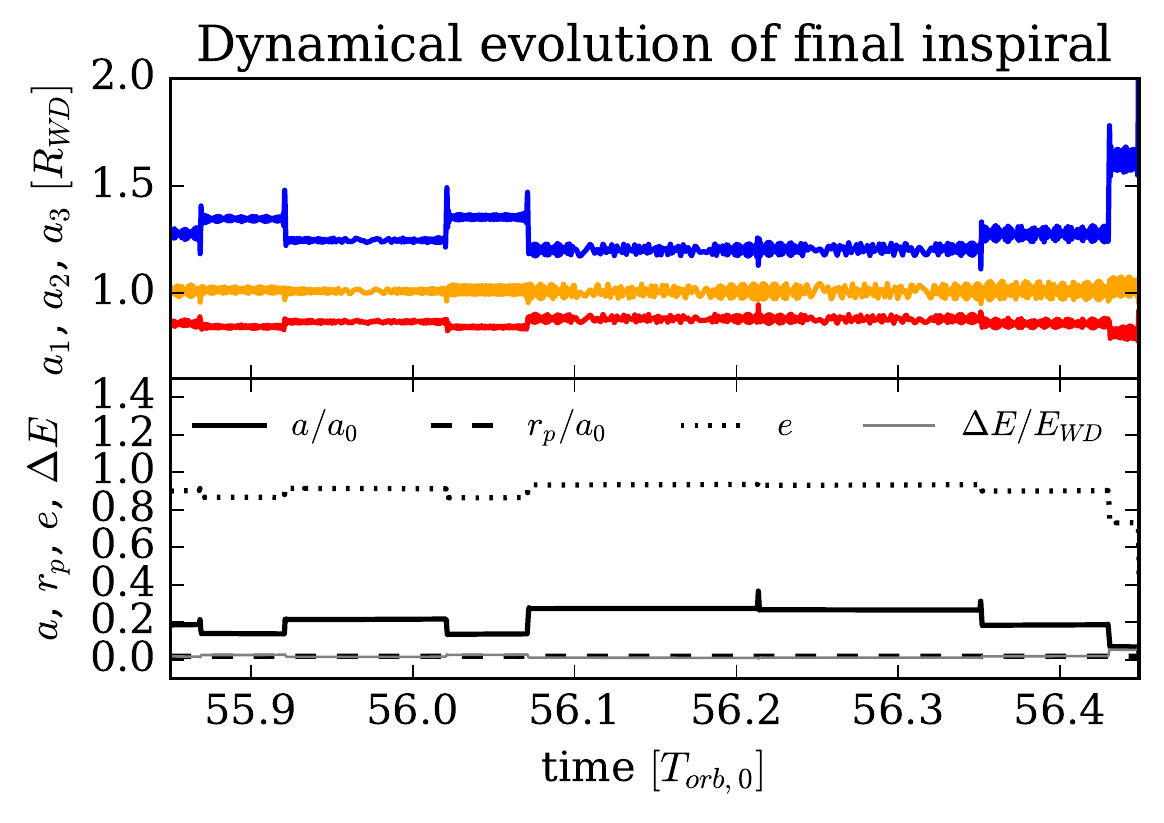}}
\caption{
Formation of a tidal inspiral between a white dwarf (WD) and a compact object (CO) during a resonant interaction resulting from an initial binary-single interaction.
The example is the same as the one shown in the top plot of Figure \ref{fig:com3body_WT_NT}.
\emph{Top:} Illustration of the binary-single interaction propagating from left to right. The colored lines show the evolution of the objects with the
blue line representing the WD. Projections of the interaction are shown with thin black lines.
\emph{Center:} Zoom in on the final few orbits of the tidal inspiral between the WD (extended object) and the CO (black dot).
The ellipsoidal shape the WD has acquired during the inspiral is here clearly seen.
\emph{Bottom:} Time evolution of the three principle axis of the WD as the WD and the CO spiral in (top window), and corresponding
orbital parameters including the energy deposited into the WD, $\Delta{E}$, in units of $E_{WD} \equiv m_{WD}^{2}/R_{WD}$ (bottom window).
The time range matches the orbits shown in the center plot. Similar inspirals form when using full hydrodynamical prescriptions \citep{2010MNRAS.406.2749L},
where the endstate is likely to be a thermonuclear transient \citep[e.g.][]{Rosswog:2009wq, 2012ApJ...746...62R}.
}
\label{fig:3d_tidalinsp_ill}
\end{figure}

\subsubsection{Monte Carlo Estimation of Cross Sections}

We follow standard prescriptions for calculating scattering
cross sections for a given outcome type $O_{i}$
by performing $N_{\rm tot}$ binary-single interactions with isotropic sampling across
a disc at infinity with radius $b$ \citep[see e.g.,][]{Hut:1983js, 2014ApJ...784...71S}. 
The corresponding 
cross section for outcome $O_{i}$ can be estimated by
\begin{equation}
\sigma_{i}= \frac{N_{i}}{N_{\rm tot}}{\pi}b^{2},
\label{numerical_cross_section}
\end{equation}
where the total number of outcomes of type $O_{i}$ from that scattering set is denoted by $N_{i}$.
The corresponding error is given by 
\begin{equation}
\Delta{\sigma_{i}}= \frac{\sqrt{N_{i}}}{N_{\rm tot}}{\pi}b^{2}.
\end{equation}
The set of interactions that did not lead to an outcome within the time limits were subsequently assumed
to have the same final outcome distribution as the set of resonant intercations.
We do not  include this correction when estimating outcome errors.

\subsubsection{Definition and Identification of Endstates}\label{sec:Definition and Identification of Endstates}

A state is identified as an \emph{exchange} if the three-body system evolves into
a binary and an unbound single where the single was  initially part of the target binary
(the initial perturber has exchanged into the target binary).
We use a binary-single tidal threshold of $0.01$ to decide if a triple state can be
labeled a binary-single state \citep[see, e.g.,][]{Fregeau:2004fj}. We do not distinguish between exchanges arising from
direct interactions (DI) and resonant interactions (RI), respectively \citep[for a definition of DI and RI the reader is refer to][]{2014ApJ...784...71S}.

We define a \emph{collision} when two objects with initial unperturbed radii $R_{1}$ and $R_{2}$ pass
each other within a distance $<(R_{1}+R_{2})$. We use the unperturbed radii even if the stars are
being tidally distorted. In this way we can directly compare the collision rates with and without tidal corrections,
which provides us with a clear estimate for how the dynamics are altered when  tidal excitations are included.
This way of defining a collision is normally known as the sticky star
approximation, and is the simplest way of including finite size effects \citep{Fregeau:2004fj}.

An \emph{inspiral} is defined here by a state composed of a binary with a SMA less than some value $a_{\rm insp} \ll a_{0}$ and a bound single.
We find that setting $a_{\rm insp} = 6(R_1 + R_2)$ results in a representative sample of inspirals. We are, however, aware that this will miss inspirals forming from pericenter passages $\gtrsim 3(R_1 + R_2)$. The  introduction of  such a threshold combined with the non-dissipative  assumption thus provides us with a strict lower limit on the inspiral cross sections. The difficulties in defining an inspiral  is associated  to our limited understanding of how a tidal inspiral dissipates energy and transfers  angular momentum into the star.

\section{Numerical Results}\label{sec: Numerical Scattering Results}

We now estimate exchange, collision, tidal, and GW inspiral cross
sections, using the numerical methods
described in Section \ref{sec: N-body with Tides and GR: Numerical Methods}.
Motivated by the scalings in Section \ref{sec:Tidal Captures from Simple Scaling Relations}, we
focus on the effect from tides as a function of the
initial semimajor axis, $a_{0}$, and the radius of the tidal object, $R$.
We study four different binary-single interactions, two of which have
one tidal object (CO-[WD-CO], CO-[MS-CO])
and two where all three objects are identical tidal objects (WD-[WD-WD], MS-[MS-MS]).
The brackets indicate what pair is in the initial binary,
where CO, WD and MS are short for compact object, white dwarf, and main sequence star,
respectively. A CO could here be a black hole (BH) or a neutron star (NS).
 For these scatterings, we use $\gamma = 4/3$ and $n=3$ for the WDs, and
$\gamma = 5/3$ and $n=3$ for the MS stars. We limit the computational time
for each scattering to $350$ initial orbital times due to limited computational resources.
Some inspirals will take more time, but those interactions are not possible to follow
at the moment. Our complementary analytical estimates, presented in Section \ref{sec:Analytical Models},
therefore serve as a crucial guideline.

To clearly isolate the effect from tides, we only consider cases
where all three objects within a given binary-single interaction have
the same mass. More physically motivated interactions, including WDs with a
realistic mass spectrum, will be explored in upcoming papers.

\subsection{Results from our N-body Scatterings}\label{sec:Results from our scattering experiments}

\begin{figure*}
\centering
{\includegraphics[width=\columnwidth]{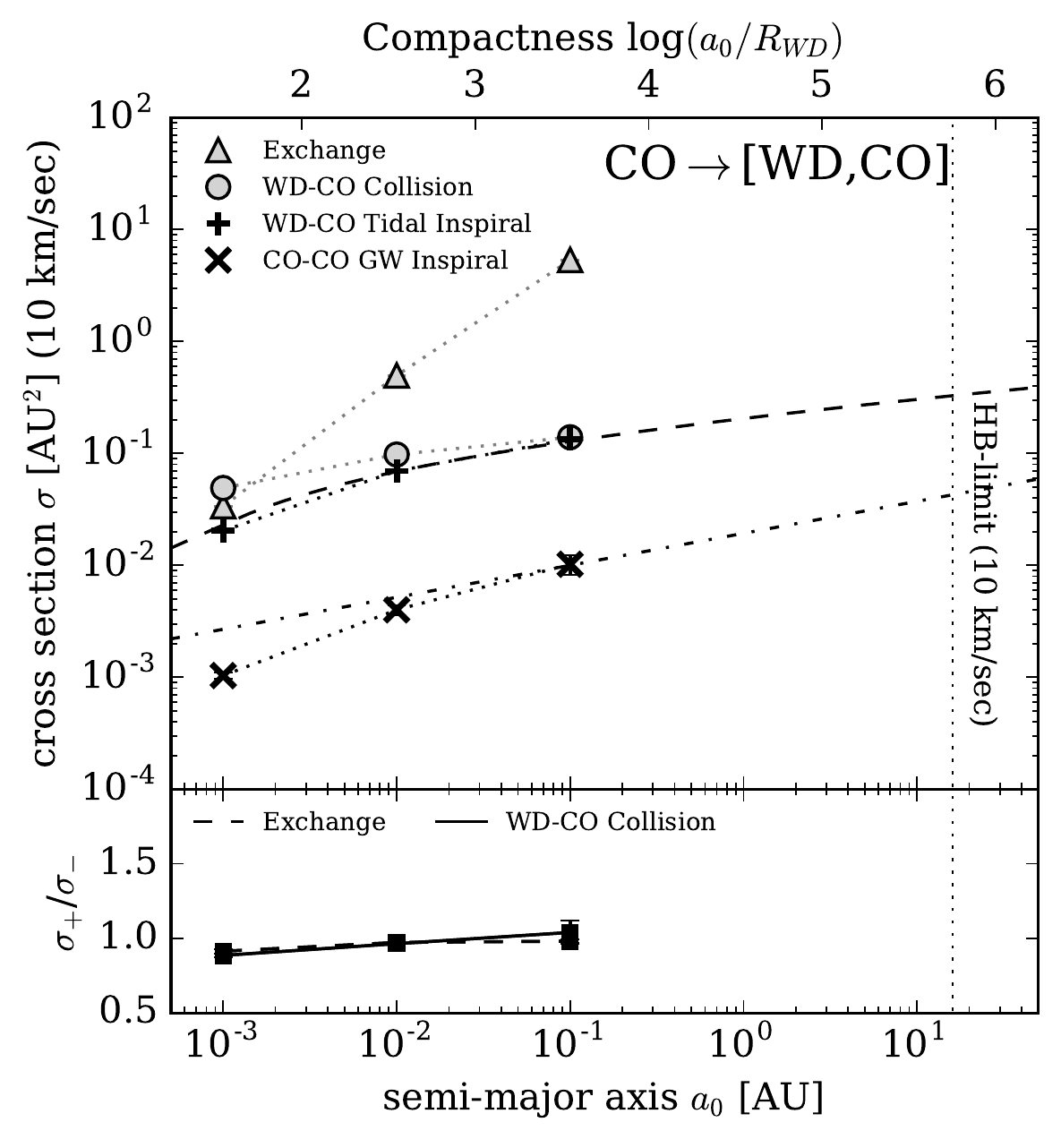}}
{\includegraphics[width=\columnwidth]{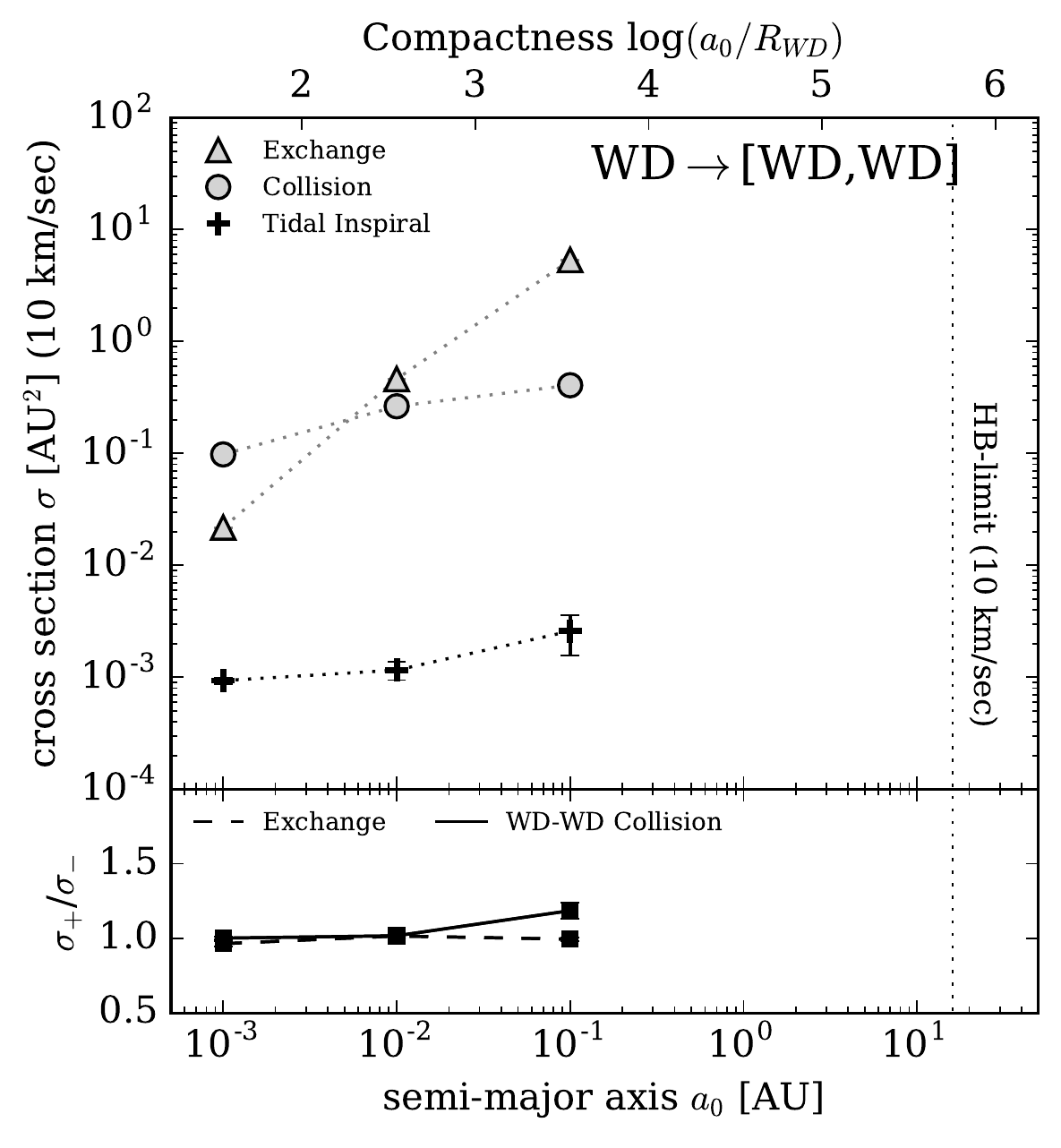}}
{\includegraphics[width=\columnwidth]{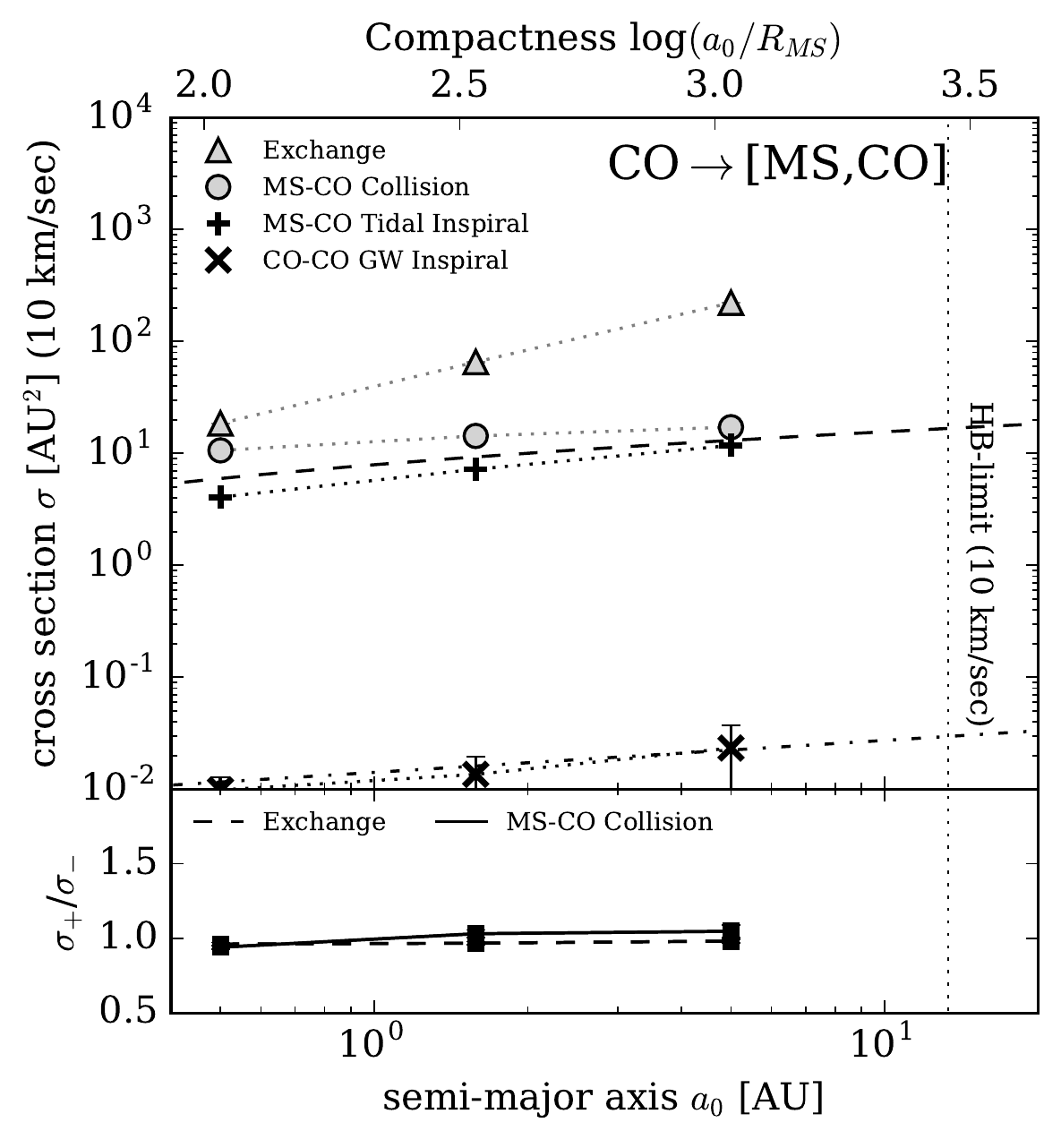}}
{\includegraphics[width=\columnwidth]{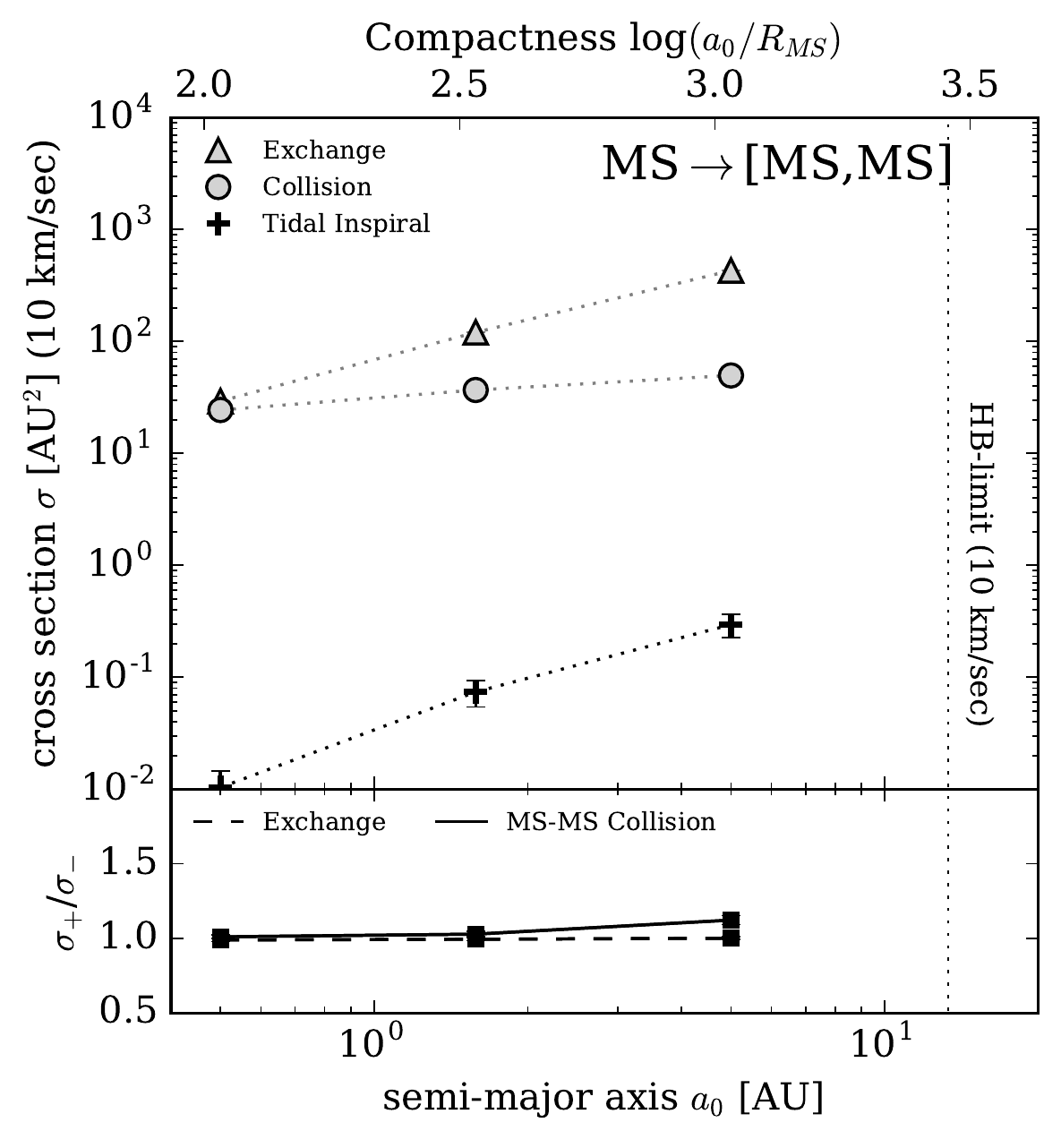}}
\caption{
Exchange, collision, tidal and GW inspiral cross sections arising from equal mass binary-single interactions,
derived using our new N-body code which includes GR and dynamical tides (see Section \ref{sec: N-body with Tides and GR: Numerical Methods}
and \ref{sec: Numerical Scattering Results}). The top window in each plot
shows the cross sections as a function of the initial SMA, $a_{0}$, of the target binary, where the bottom window shows the ratio between
cross sections derived with $(\sigma_{+})$ and without $(\sigma_{-})$ tides included in the EOM. The x-axis at the top of each window
shows the SMA scaled by $R$, where $R$ is the physical radius of the extended tidal object in the triple interaction.
As we show in Section \ref{sec:Analytical Models}, the
combination $a_{0}/R$ is the relevant parameter for determining the rate of inspirals relative to collisions.
The vertical dotted lines show the HB limit for each system assuming $v_{\infty}=10\ \text{km}\ \text{s}^{-1}$:
the cross sections are therefore only valid to the left of these lines for this value of $v_{\infty}$. The dashed and dashed-dotted lines
illustrate our analytical scaling solutions for the tidal and GW inspiral cross sections, respectively, as described in
Section \ref{sec:Analytical Models} and summarized in Figure \ref{fig:crosssec_summfig}. 
The scattering results shown here are discussed in Section \ref{sec:Results from our scattering experiments}.
}
\label{fig:cs_all}
\end{figure*}

The scattering results from our $3$-body experiments are shown in Figure \ref{fig:cs_all}. The upper window in each
plot shows the cross sections with tides and GR included in the EOM,
where the lower window shows the ratio between cross sections calculated with ($\sigma_{+}$) and
without ($\sigma_{-}$) tides. The bottom $x$-axis shows $a_{0}$ in AU, while the top shows the
compactness $a_{0}/R$. The upper limit of $a_{0}/R \approx 3-4 $ is set by computational limitations.
The derived cross sections are based on $5\times10^4$ scatterings per SMA, $a_{0}$.  
We describe our specific  findings below.

\subsubsection{CO-[WD-CO]}\label{sec:WDNSNS}
We first study the binary-single interaction between a CO($1.2 M_{\odot}$) and
a [WD($1.2 M_{\odot}$, $0.006 R_{\odot}$)-CO($1.2 M_{\odot}$)] binary. Results are shown in the upper
left window in Figure \ref{fig:cs_all}. The inclusion of tides clearly results in a
population of WD-CO tidal inspirals with a cross section
that increases with $a_{0}$. As seen, our analytical estimate (dashed line) from Section
\ref{sec:Analytical Models} indicates that the
inspiral rate will exceed the classical collision rate for $a_{0} \gtrsim 0.2$ AU.
In the bottom window we see that the inclusion of tides doesn't affect either the exchange or collision cross sections
in a significant way.

\subsubsection{CO-[MS-CO]}
The WD from our previous  study  is now replaced by
a MS star ($1.0 M_{\odot}$, $1.0R_{\odot}$), and the two COs are each given a mass of $1.0 M_{\odot}$. Results are shown in the bottom
left window in Figure \ref{fig:cs_all}. Tidal inspirals form in this case,
but the rate relative to the collision rate is now significantly lower when compared
to the HB limit (vertical dotted line). As a result, MS-CO tidal inspirals do not
have the necessary range in $a_{0}$ to clearly dominate over collisions.
From the results in the bottom window, we also in this case conclude that tides
do not significantly affect the classical rate of exchanges and collisions.

\subsubsection{WD-[WD-WD]}
We now consider the interaction between three WDs ($1.2 M_{\odot}$, $0.006 R_{\odot}$).
Results are shown in the upper right window in Figure \ref{fig:cs_all}. Tidal inspirals
form but at a very low rate compared to when one of the objects is a point mass.
WD-WD tidal inspirals seem therefore not to contribute significantly
to the WD coalescence rate at any $a_{0}$, even at the hard binary limit. More extensive
numerical simulations must be performed to investigate this further.
Again, we see in the bottom window that the effect from tides does not significantly alter
either the exchange or collision cross sections.

\subsubsection{MS-[MS-MS]}
As a final example, we study the
interaction between three MS stars ($1.0 M_{\odot}$, $1.0R_{\odot}$).
Results are shown in the bottom right window in Figure \ref{fig:cs_all}. 
The rate of inspirals relative to collisions is even lower than in the WD-[WD-WD] case,
which results in an inspiral rate that is about 1-2 orders of magnitude
lower than the collision rate at the hard binary limit. 
As in the other cases, no strong effects are seen from the inclusion of
tides on the exchange and collision cross sections in the lower window.

\subsection{Summary of our N-body Scatterings}

Our numerical results indicate
that tides do not strongly affect the classical exchange and collision cross sections.
Instead, when tides are included we see a clear population of
tidal inspirals appearing, as initially speculated by \cite{1992ApJ...385..604K}.
For the first time we estimate here the cross section and we find that it
increases with $a_{0}$, as opposed to the classical sticky star collision
cross section \citep{Fregeau:2004fj} which stays nearly flat (the analytical solution also gives a constant  cross section). Tidal inspirals therefore have
the possibility to dominate over collisions. Because the two stars undergoing the tidal inspiral are likely to merge \citep{1993PASP..105..973R, 2010MNRAS.406.2749L},
tidal inspirals can actually dominate the coalescence rate for some interaction channels.

The scattering results further suggest that the rate of tidal inspirals relative to collisions
depends on the compactness $a_{0}/R$ and not only $a_{0}$.
For example, if we compare the two datasets CO-[WD-CO] and CO-[MS-CO], we see that
the inspiral and collision cross sections are equal at the same $\log(a_{0}/R)\approx3.5$, but not at a given value of $a_{0}$.
This is consistent with the simple scalings from Section \ref{sec:Tidal Captures from Simple Scaling Relations}, which depend on $a_{0}/R$.
The maximum number of inspirals relative to collisions is therefore set by how large $a_{0}/R$ can be within the HB limit.
We know the HB limit scales with the mass of the objects as $a_{\rm HB} \propto m/v_{\infty}^2$
with no dependence on $R$ (see Equation \ref{eq:v_c}). As a result, the smaller
an object is compared to its mass the larger number of inspirals relative to collisions can be generated during binary-single encounters.
This explains why the CO-[WD-CO] channel  produces the most inspirals relative to collisions in the scattering examples presented here. 

Gravitational wave inspirals between CO-CO binaries are also shown for the WD and MS datasets.
The estimated cross sections are fully consistent with the results from \cite{2014ApJ...784...71S}, which
provides further credence in our numerical methods. In \cite{2014ApJ...784...71S} it was shown
that these GW inspirals are likely to be the predominating channel for high eccentricity NS-NS GW mergers detectable by LIGO.

Finally, we see that the rate of tidal inspirals between two tidal objects is significantly
lower than the rate involving a tidal object and a compact object.
There are at least two physical reasons for this.
First, an IMS binary with pericenter $R<r_{\rm p}<2R$ will  lead to a collision instead of an inspiral when both objects are extended.
Second, the energy deposited into tides falls off very steeply with distance $r$ (about $r^{-9}$ in the PT model) which here greatly suppresses
the formation of inspirals which now need to arise from passages
with $r_{\rm p} > 2R$. Our imposed tidal threshold described in Section \ref{sec:Definition and Identification of Endstates}
also plays a role; their could very well be inspirals forming through weak interactions evolving over hundred to thousand of orbits
that we are not able to follow and identify. 

In this work we have considered high mass
WDs which have a relative stiff equation of state ($n=3$), the inspiral cross section could very well be much higher
for more realistic low mass WDs which are more prone to tidal deformations due to their lower polytropic index ($n=1.5$). This motivates our future work on unequal mass interactions and, in particular, those  involving WDs and COs.
 
In the following section we present an analytical model that can
explain all the main trends we have seen in our simulations  so far.
Our numerical scattering results are discussed further in the context of this model in Section \ref{sec:Discussion}.

\section{Analytical Description of Inspirals and Collisions}\label{sec:Analytical Models}

In this section we present an analytical model for describing the inspiral cross section and how it
depends on the properties of the interacting objects and the initial orbital parameters of the binary-single system.
In contrast to the simple scalings from Section \ref{sec:Tidal Captures from Simple Scaling Relations},
we now correctly account for the energy loss arising from multiple close passages during the binary-single interaction.
For all calculations and results we assume the equal mass case,
and for tidal inspirals and collisions we further assume the IMS binary is composed
of one point-mass perturber and one tidal object with radius $R$.
Further details on the approach we here use to calculate the inspiral cross section can be found in \cite{2014ApJ...784...71S}.

\subsection{Energy Losses and Orbital Evolution}
We start by considering the orbital energy evolution of an IMS binary with initial eccentricity $e$, semimajor axis $a$,
and corresponding pericenter $r_{\rm p} = a(1-e)$.
We assume the binary loses a constant amount of orbital energy $\Delta{E}_{\rm p}$ at each pericenter passage due to
some effect which only depends on $r_{\rm p}$. This is a reasonable approximation for tides and GR, which
both have a very steep dependence on $r_{\rm p}$ \citep[for tides and GR see, e.g.,][respectively]{1977ApJ...213..183P, Blanchet:2006kp}.
The corresponding angular momentum loss $\Delta{L}_{\rm p}$
is generally small compared to the energy loss (see discussion from Section \ref{sec:Tidal Capture Example}), $r_{\rm p}$ and $\Delta{E}_{\rm p}$ will therefore not
change significantly until the binary circularizes. This is also seen in Figure \ref{fig:twobody_tidalcapture_NSMS}.
As a result, the orbit averaged energy evolution can be written as,
\begin{equation}
\frac{dE}{dt} \approx \frac{\Delta{E}_{\rm p}}{T_{\rm orb}(t)} \approx \frac{2\Delta{E}_{\rm p}}{\pi m^{5/2}} E(t)^{3/2},
\end{equation}
where $t$ is time, $m$ is the mass of one of the (equal mass) objects, ${T_{\rm orb}(t)}$ is the orbital time of the IMS binary, and $E(t)$ is the corresponding orbital
energy. Using this relation, we find the solution for the time evolving SMA $a(t) = m^{2}/[2E(t)]$ of the IMS binary to be given by,
\begin{equation}
a(t) = a \left( 1 - t \frac{ \Delta{E}_{\rm p}}{{\pi}\sqrt{2am^{3}}} \right)^{2}.
\label{eq:DEoE0}
\end{equation}
We see that the two IMS binary members will merge in a finite time, which corresponds to
the limit where $a(t) \rightarrow 0$. This time we define as the inspiral time, $t_{\text{insp}}$, and is from 
Equation \eqref{eq:DEoE0} found to be,
\begin{equation}
t_{\text{insp}} = \pi \sqrt{2} \frac{m^{3/2}\sqrt{a}}{\Delta{E}_{\rm p}}.
\label{eq:t_insp}
\end{equation}
For an IMS binary to undergo a successful inspiral, its inspiral time must be less than the time it is isolated from the bound single.
Following \cite{2014ApJ...784...71S}, the time that the IMS binary is isolated from the single
is given by,
\begin{equation}
t_{\rm iso} = 2\pi \sqrt{\frac{a_{\rm bs}^3}{3m}},
\label{eq:t_iso}
\end{equation}
where $a_{\rm bs}$ is the semimajor axis of the bound single relative to the IMS binary.
This semimajor axis
can be found from energy conservation assuming the total orbital energy has not changed before the formation of the IMS binary,
\begin{equation}
E_{0} = \frac{m^2}{2a_{0}}=\frac{m^2}{2a} + \frac{2m^2}{2a_{\rm bs}}.
\end{equation}
By solving for $a_{\rm bs}$ in the above equation
we find $a_{\rm bs} = 2a_{0}/(1-1/a')$, where $a' \equiv a/a_{0}$. Using this expression for $a_{\rm bs}$, the isolation time $t_{\rm iso}$
from Equation \eqref{eq:t_iso} can now be written as
\begin{equation}
t_{\rm iso} =  \frac{4}{\sqrt{3}}\left( \frac{a'}{a'-1} \right)^{3/2}  2\pi \sqrt{\frac{a_{0}^3}{2m}},
\end{equation}
where the last part equals the orbital time of the initial binary.

While the form of $\Delta{E}_{\rm p}$ has so far been left general, to proceed we need to
specify a functional form for $\Delta{E}_{\rm p}$. In the sections below we derive analytical cross sections for an energy
loss term of the form $\Delta{E}_{\rm p} \propto r_{\rm p}^{-\beta}$.

\begin{figure}
\centering
\includegraphics[width=\columnwidth]{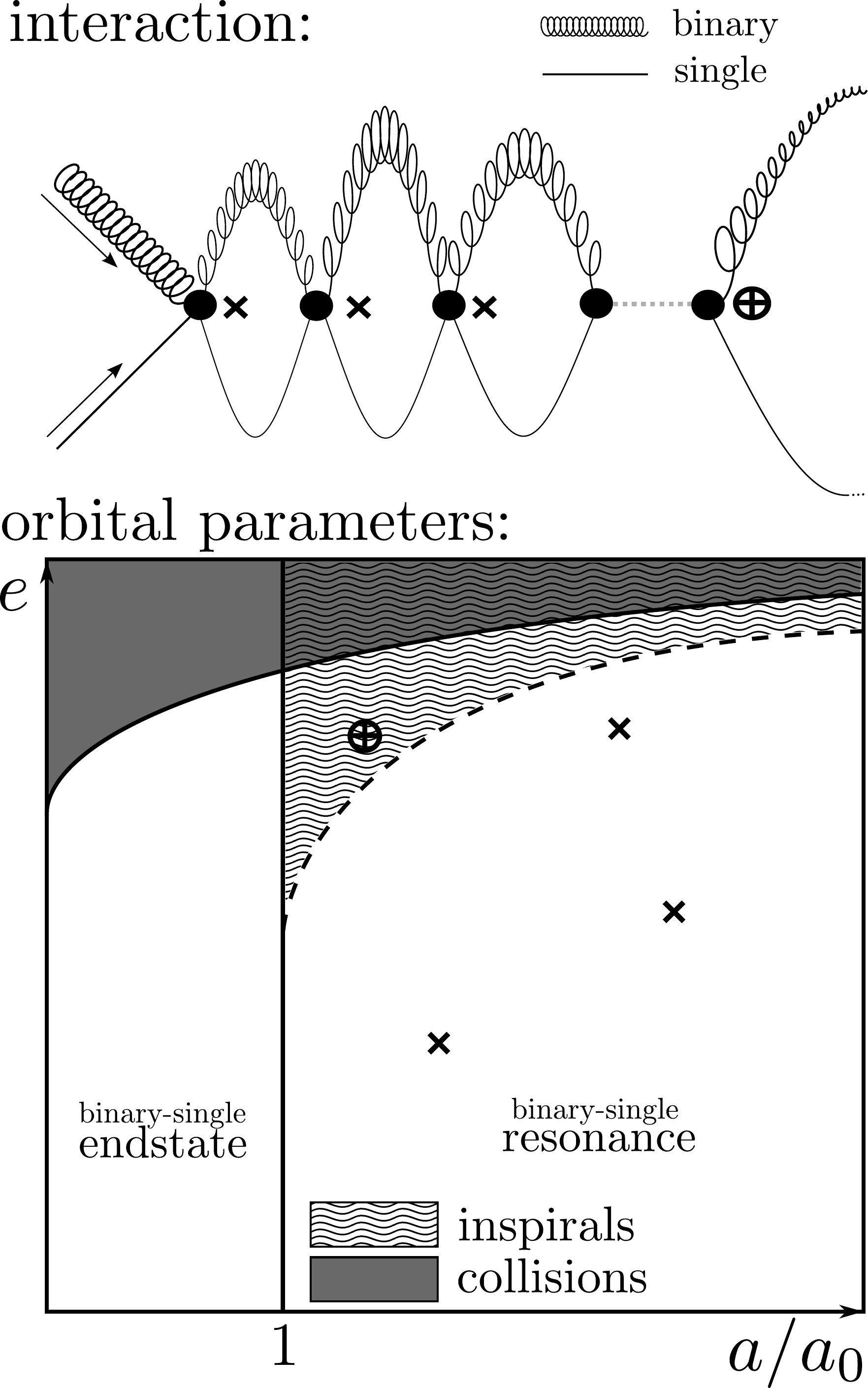}
\caption{Illustration of a resonant binary-single interaction in real space (top) and orbital phase space (bottom), with regions
highlighting where inpirals and collisions can form.
\emph{Top}: Illustration of a resonant binary-single interaction
evolving towards its endstate, which is here an inspiral, from {\it left} to {\it right}.
As illustrated, a resonant system often evolves through a series of intermediate binary-single states characterized by an IMS binary with a bound single.
Each time all three objects come together (large black dots), the objects mix and each IMS binary can therefore
consist of any two of the three objects. A general evolution can consist of many such intermediate binary-single states, which are here denoted by
a series of small grey dots. The IMS binaries are formed with a wide distribution in semimajor
axis $a$ and eccentricity $e$. If the corresponding pericenter distance is small enough, GW radiation or tidal effects
will lead to a significant orbital energy loss at pericenter. As a result, the IMS binary will quickly
spiral in with a possible merger to follow -- similar to GW or
tidal captures in the field.
\emph{Bottom}: Orbital phase space spanned by $a/a_{0}$, where $a_{0}$ is the initial semimajor axis, and $e$ is the eccentricity.
Each IMS binary is formed somewhere in this space (a few examples which match with the illustration at the top
are shown by the small thick symbols) with a non-negligible
probability for either be formed in the inspiral region (wavy part) or collision region (solid grey).
The areas of the two regions scale differently with $a_{0}$, which
makes collisions dominant at low $a_{0}$ and inspirals at high $a_{0}$ (see Section \ref{sec:Inspiral Mergers Relative to Collisions}). 
If a binary is formed to the left of $a/a_{0} = 1$, a classical binary-unbound-single endstate
has been formed, such as an exchange.
Section \ref{sec:Analytical Models} explains this in detail.
}
\label{fig:ae_ill}
\end{figure}

\subsection{Energy Loss Term of the form $\Delta{E} \propto r_{\rm p}^{-\beta}$}

We consider the following generic form for the
orbital energy loss at pericenter, 

\begin{equation}
\Delta{E}_{\rm p} = \mathcal{E} \frac{Gm^2}{\mathcal{R}} \left(\frac{\mathcal{R}}{r_{\rm p}}\right)^{\beta},
\label{eq:general_E_rp}
\end{equation}
where $\mathcal{E}$ is a normalization factor, and $\mathcal{R}$ is a characteristic radius.
This form represents GR and tidal effects reasonably well, as now will be described.

\subsubsection{Energy Losses from Gravitational Waves}

For GR the leading order dissipative term is quadrupole GW radiation \citep{Peters:1964bc}, which in the high eccentric equal mass case
leads to the following orbital energy loss at each pericenter passage \citep{Hansen:1972il},
\begin{equation}
\Delta{E}_{\rm GW}	\approx \frac{85\pi}{12}\frac{G^{7/2}}{c^{5}}\frac{m^{9/2}}{{r_{\rm p}}^{7/2}}.
\label{eq:deltaE_GR}
\end{equation}
By a simple rearrangement one finds the above formulation for $\Delta{E}_{\rm GW}$ can be
expressed equally by Equation \eqref{eq:general_E_rp}, by setting
\begin{equation}
\beta = \frac{7}{2},\ \mathcal{E} = \frac{85\pi\sqrt{2}}{96},\ \mathcal{R} = R_{s} = \frac{2Gm}{c^{2}},
\end{equation}
where $R_{s}$ is here the Schwarzschild radius of an object with mass $m$.

\subsubsection{Energy Losses from Stellar Tides}

For tides, the energy deposited into tidal mode oscillations during a single pericenter passage
is to leading order ($l=2$ quadrupole tides) in the PT formalism given by,
\begin{equation}
\Delta{E}_{\rm tid}	\approx \frac{Gm^{2}}{R} \left(\frac{R}{r_{\rm p}}\right)^{6}T_{2}(\eta),
\label{eq:deltaE_tid}
\end{equation}
where $R$ is the stellar radius, and $T_{2}(\eta)$ is a non-trivial function which depends on the stellar structure
and the dimensionless parameter $\eta$, which in the equal mass case equals
\begin{equation}
\eta = \frac{1}{\sqrt{2}} \left( \frac{R}{r_{\rm p}} \right)^{-3/2}.
\end{equation}
For our analytical model we approximate $T_{2}(\eta)$
by a simple powerlaw of the form,
\begin{equation}
T_{2}(\eta) \approx A\eta^{-\alpha}.
\end{equation}
This approximation was also used in, e.g., \cite{1986ApJ...306..552M} and \cite{1993ApJ...412..593L}. 
From combining the terms we see that in the case of tides, the form for $\Delta{E}_{\rm tid}$ can be expressed by Equation \eqref{eq:general_E_rp}
with
\begin{equation}
\beta = 6+\frac{3\alpha}{2},\ \mathcal{E} = A\sqrt{2^{\alpha}},\ \mathcal{R} = R.
\end{equation}
For our analytical estimations in this work we use $\beta=9$ ($\alpha=2$), which
is applicable for the $n=3$ polytropic examples we study here.

\subsubsection{Resultant Inspiral Time}

With the general form for $\Delta{E}_{\rm p}$ given by Equation \eqref{eq:general_E_rp}, one can now write the inspiral time $t_{\text{insp}}$ from
Equation \eqref{eq:t_insp} as,
\begin{equation}
t_{\text{insp}} = 2\pi r_{\rm p}^{\beta} \sqrt{a} \frac{\mathcal{R}^{1-\beta}}{\mathcal{E}\sqrt{2m}}.
\end{equation}
In the following section we compare this time with the isolation time to find which IMS binaries that are able to
undergo an inspiral.

\subsection{Formation of Inspirals in Orbital Phase Space}\label{sec:Formation of Inspirals in orbit space}

An IMS binary is able to undergo an inspiral if its inspiral time is less than the isolation time.
The set of IMS binaries that are able to inspiral can therefore be found by first finding the combination of
$a'$ and $e$ which fulfills $t_{\text{insp}} = t_{\text{iso}}$. In this case we find the following relation,
\begin{equation}
\epsilon_{\rm insp} = \mathcal{E}^{1/\beta} \left( {a_{0}}/{\mathcal{R}} \right)^{1/\beta - 1} \left[ \frac{4}{\sqrt{3}} \frac{a'}{{a'}^{\beta} (a' - 1)^{3/2}} \right]^{1/\beta},
\label{eq:e_insp}
\end{equation}
where $\epsilon_{\rm insp} \equiv 1-e_{\rm insp}$, and $e_{\rm insp}$ is the eccentricity. This formula relates the semimajor axis, $a'$, of an IMS binary
to the eccentricity, $e_{\rm insp}$, the binary must have to exactly spiral in before the single returns. If the eccentricity is larger than
$e_{\rm insp}$ (closer to $1$), the pericenter is smaller, which results in a faster inspiral due to the increased energy loss at pericenter.
Equation \eqref{eq:e_insp} therefore defines the boundary of the region in orbital phase space (spanned by $a,e$) in which inspirals can form ($\epsilon < \epsilon_{\rm insp}$).
This is shown in Figure \ref{fig:ae_ill} and further explained in the corresponding caption.

\subsection{Inspiral Cross Section}\label{sec:Inspiral cross sections}

The cross section for an outcome $x$ arising from a RI,
can be factorized in the following way \citep{2014ApJ...784...71S},
\begin{equation}
\sigma_{\rm x} = \sigma_{\rm RI} \times P(x | {\rm RI}),
\label{eq:sigma_x}
\end{equation}
where $\sigma_{\rm RI}$ is the cross section for a binary-single interaction to evolve as a RI, and $P(x | {\rm RI})$ is the
probability for $x$ to be an endstate given the interaction is a RI.
To estimate the inspiral cross section we must therefore
calculate the probability for an inspiral to form during
a resonant interaction, $P({\rm insp} | {\rm RI})$. This term is proportional to the probability for an IMS binary
to form with parameters $a',e$ inside the inspiral region (see Section \ref{sec:Formation of Inspirals in orbit space} above) doing a resonance.
In \cite{2014ApJ...784...71S} it was shown that the distribution in $a'$ and $e$ sampled in a resonance is approximately
flat at high eccentricity, as a result, the relative probability for a binary to form in some high eccentricity region scales with
the area of that region.
The majority of inspirals have a very
high eccentricity \citep{2014ApJ...784...71S}, therefore the probability for forming an inspiral is proportional to the area of the inspiral region.
This area is found by integrating $\epsilon_{\rm insp}$ from Equation \eqref{eq:e_insp} over $a'$ from $a' \approx 1$
to $a' \approx 2$ (when $a' \gtrsim 2$ the triple system can no longer be considered as a binary with a bound single).
The $a'$ dependent term in the brackets from Equation \eqref{eq:e_insp} integrates to a constant,
so the area, and thereby the inspiral probability, will simply scale as,
\begin{equation}
P({\rm insp} | {\rm RI}) \propto \mathcal{E}^{1/\beta} \left( {a_{0}}/{\mathcal{R}} \right)^{1/\beta - 1}.
\label{eq:P_insp}
\end{equation}
As a result, the inspiral probability, has dependence only on the strength of the loss term $\mathcal{E}$,
its slope $\beta$, and the compactness of the initial binary $(a_{0}/\mathcal{R})$.
In the HB limit $\sigma_{\rm RI}$ is proportional to the CI cross section $\sigma_{\rm CI}$
from Equation \eqref{eq:sigma_CI}.
Writing out Equation \eqref{eq:sigma_x} for inspirals we now finally find the inspiral cross section to scale as,
\begin{equation}
\sigma_{\rm insp} \propto   \frac{m \mathcal{R}}{v^{2}_{\infty}} \left[ \mathcal{E}^{1/\beta} \left(\frac{a_{0}}{\mathcal{R}}\right)^{1/\beta} \right]
\label{eq:sigma_insp}
\end{equation}
From this relation we can conclude that
the cross section for any kind of inspiral \emph{always increases} with $a_{0}$ (since $\beta$ is always positive).
The rate of inspirals resulting from any pericenter-dependent loss term is therefore dominated by widely separated
binaries and not tight binaries, as one might naively guess. 
This was illustrated for GW inspirals in \cite{2014ApJ...784...71S}, however, here we have generalized the framework to
show that this actually is a generic feature of any kind of inspiral, including tidal inspirals.

\subsection{Collision Cross Section}\label{sec:collisions}

The collision cross section, $\sigma_{\rm coll}$, can be estimated by a similar approach
as the one described for inspirals. As described in Section \ref{sec:Definition and Identification of Endstates}, we define a
collision to be when two objects pass
each other at a distance smaller than their total unperturbed radii without
inspiraling first. In the case of an IMS binary composed of a point-mass perturber and
a tidal object with radius $R$, a collision will therefore occur if the pericenter distance, $r_{\rm p}$, is smaller than $R$.
To estimate the cross section for such as collision we first need to
calculate the minimum eccentricity, $e_{\rm coll}$, a temporary formed binary
with semimajor axis $a'$ must have to collide.
For this we use the standard relation $r_{\rm p} = a_{0}a'(1-e)$ and substitute $r_{\rm p}$ with $R$, from which
we now find,
\begin{equation}
\epsilon_{\rm coll} = \frac{R}{a_{0}} \frac{1}{a'},
\label{eq:epsilon_coll}
\end{equation}
where $\epsilon_{\rm coll} \equiv 1-e_{\rm coll}$.
This relation defines the boundary of the collision region in
orbital phase space illustrated in Figure \ref{fig:ae_ill}.
As for the inspirals, integrating $\epsilon_{\rm coll}$ over $a'$ lead us to the relevant scaling for the collision
probability given a RI,
\begin{equation}
P({\rm coll} | {\rm RI}) \propto {R}/{a_{0}}.
\end{equation}
This can now be converted into a cross section using Equation \eqref{eq:sigma_x},
\begin{equation}
\sigma_{\rm coll} \propto \frac{mR}{v^{2}_{\infty}}.
\label{eq:sigma_coll}
\end{equation}
The collision cross section is therefore independent of $a_{0}$ and linear in $R$.

By comparing our analytical expressions for the tidal inspiral cross section (in which case $\mathcal{R}$
should be replaced by $R$) and the collision cross section we
observe a few interesting similarities. First,  we see that the
tidal inspiral cross section approaches the collision cross section as $\beta \rightarrow \infty$.
In terms of cross sections, our simple $\beta$-model from Equation \eqref{eq:general_E_rp}
therefore seems to be the appropriate leading order extension for describing effects related to finite sizes,
including non-dissipative solid-sphere collisions. Second, we notice that the inspiral cross section is similar to the
collision cross section if the star is treated as a solid sphere with radius $\propto R(a_{0}/R)^{1/\beta}$
instead of just $R$. The relevant radius of the star is therefore not just a constant times its radius -- it further includes
a factor that scales with the energy of the few-body system it evolves in. This was also noticed by our simple scalings
in Section \ref{sec:Tidal Captures from Simple Scaling Relations}.

\subsection{Inspirals in the Collision Dominated Regime}\label{sec:Inspirals in the collision dominated limit}

The inspiral and collision regions in orbit space overlap as illustrated in Figure \ref{fig:ae_ill}, which means that
IMS binaries with high enough eccentricity will collide instead of spiraling in. We did not take this into account
when calculating the inspiral cross sections in Section \ref{sec:Inspiral cross sections}.
With the understanding of where collisions form from Section \ref{sec:collisions}, we can now correct for
this overlap. We only write out the solution for tidal inspirals, since the correction is never really important
for GW inspirals -- we therefore replace $\mathcal{R}$ with R below.

The asymptotic inspiral solution given by Equation \eqref{eq:sigma_insp} assumed
that inspirals can form in the full inspiral area, we can therefore write the collision corrected
solution, here denoted by $\sigma_{\rm insp-c}$, as a product of the asymptotic solution where collisions play no role, $\sigma_{\rm insp}$,
and a weight term specifying the fraction of the full inspiral area that is not overlapping with the collision area.
In Figure \ref{fig:ae_ill} this is the wavy region between the dashed and the solid line.
The collision corrected inspiral cross section can therefore be written as, 
\begin{equation}
\sigma_{\rm insp-c} \approx \sigma_{\rm insp}\left[ \frac{\int_1^{a'_{\rm ic}} (\epsilon_{\rm insp}-\epsilon_{\rm coll})da'}{\int_1^{a'_{\rm u}} \epsilon_{\rm insp} da'}\right],
\label{eq:sigma_cc_insp_1}
\end{equation}
where $a'_{\rm ic}$ is where $\epsilon_{\rm insp}$ crosses $\epsilon_{\rm coll}$,
and $a'_{\rm u}$ is the maximum value for $a'$ ($a'_{\rm u} \approx 2$).
Assuming we know the normalizations of $\sigma_{\rm coll}$ and $\sigma_{\rm insp}$, one can now solve for the
full collision corrected inspiral cross section, $\sigma_{\rm insp-c}$. This will be done using numerical techniques
in the next section. However, even without normalizations, we can estimate how $\sigma_{\rm insp-c}$
scales with $a_{0}/R$ in the limit where collisions dominate. In this regime we know that $a'_{\rm ic}$ must be close to $1$
in which case $a'_{\rm ic}$ to leading order can be written as,
\begin{equation}
a'_{\rm ic} \approx 1+\left(\frac{4\mathcal{E}}{{\sqrt{3}}}\frac{a_{0}}{R}\right)^{2/3}.
\label{eq:ap_upper}
\end{equation}
In this limit the integrals in Equation \eqref{eq:sigma_cc_insp_1} can also be solved
by Taylor expanding around $\delta=0$, where we here define for convenience $\delta \equiv a'_{\rm ic}-1$. 
As a result, the $a'^{1/\beta - 1}$ term in $\epsilon_{\rm insp}$ can be dropped which leaves us with the term $(a'-1)^{-3/(2\beta)}$ and $\epsilon_{\rm insp}$ can now
be trivially integrated.
The integral over  $\epsilon_{\rm coll}$ is also easily found and will scale $\propto \text{ln}(1+\delta)$, which is $\approx \delta$ when $\delta \ll 1$. 
By writing out the full expression in Equation \eqref{eq:sigma_cc_insp_1} following these assumptions,
we find the scaling $\sigma_{\rm insp-c} \propto a_{0}^{2/3}$, which holds in the low $a_{0}/R$ limit.

To summarize, the inspiral cross section scales differently with $a_{0}$ depending on if collisions are dominating $(a_{0}/R \rightarrow 1)$
or not $(a_{0}/R \rightarrow \infty)$, with specific scaling solutions given by,
\begin{equation}
\sigma_{\rm insp-c} \propto a_{0}^{1/\beta},\ (a_{0}/R) \rightarrow \infty 
\end{equation}

\begin{equation}
\sigma_{\rm insp-c} \propto a_{0}^{2/3},\ (a_{0}/R) \rightarrow 1.
\end{equation}
Cross section results from a full integration of Equation \eqref{eq:sigma_cc_insp_1} including
correct normalizations are shown in Figure \ref{fig:crosssec_summfig}, and will be discussed in the following
section.

\subsection{Numerical Calibration of Analytical Cross Sections}\label{sec:Numerical Calibration of Analytical Cross Sections}
We here show the analytical cross sections with correct
normalizations estimated using our numerical simulations from Section \ref{sec: Numerical Scattering Results} with tides and GR.
As for the analytical results, the scalings presented in this section
are only valid in the equal mass case.
The cross sections are given in the following rescaled form for convenience, 
\begin{equation}
\bar{\sigma} \equiv \sigma \left[ \frac{(m/M_{\odot})(\mathcal{R}/R_{\odot})}{(v_{\infty}/\text{km}\ \text{s}^{-1})^{2}}\right]^{-1}.
\end{equation}
Since the collision corrected tidal inspiral cross section from Section \ref{sec:Inspirals in the collision dominated limit}
has no closed form across the full interval in $a_{0}/R$, we instead present tidal inspirals plus collisions which, in our model,
scales as $(a_{0}/R)^{1/9}$ for $\text{log}(a_{0}/R)>1$ (see Figure \ref{fig:crosssec_summfig}). 

\subsubsection{Cross Sections}

The analytical cross section for a tidal extended object and a point-mass perturber
to undergo a tidal inspiral \emph{or} a collision (the total coalescence rate) is given by,
\begin{equation}
\bar{\sigma}_{\rm insp,tid} + \bar{\sigma}_{\rm coll} \approx 727 \left( \frac{a_{0}}{R} \right)^{1/9} \ \text{AU}^{2},
\label{eq:sigma_insp_p}
\end{equation}
for $\text{log}(a_{0}/R)>1$. The normalization is here valid for polytropes with index $n=3$ (the exact value for $\gamma$
do not play a significant role here).
The cross section for two compact objects to undergo
a GW inspiral is,
\begin{equation}
\bar{\sigma}_{\rm insp,GW} \approx 2095 \left( \frac{a_{0}}{R_{s}} \right)^{2/7} \ \text{AU}^{2}.
\label{eq:sigma_insp_GW}
\end{equation}
A compact object is here either a NS or a BH -- a WD is not compact enough
for GWs to dominate over tides during close encounters.
For a collision between an extended tidal object and a point-mass perturber we find,
\begin{equation}
\bar{\sigma}_{\rm coll} \approx  924 \ \text{AU}^{2}.
\label{eq:sigma_coll_p}
\end{equation}
A few examples are given below.

\subsubsection{Examples}

To illustrate how to use the scaling relations from above, let us now consider three examples related to the
binary-single interaction between a NS($1.2M_{\odot}$)
and a [WD($1.2M_{\odot}, 0.006R_{\odot}$)-NS($1.2M_{\odot}$)] binary with
$a_{0}=5\text{AU} \approx 1075R_{\odot}$ in a cluster with $v_{\infty}=10\ \text{km}\ \text{s}^{-1}$.

\begin{itemize}
\item \emph{Tidal inspirals +  collisions:} The total WD-NS coalescence cross section (tidal inspirals + collisions)
can be estimated using Equation \eqref{eq:sigma_insp_p} from which we find
${\sigma}_{\rm insp,tid} + {\sigma}_{\rm coll} \approx 2\cdot[1.2\cdot0.006/10^2]\cdot727\cdot(1075.0/0.006)^{1/9}\ \text{AU}^2 \approx 0.4\ \text{AU}^2$.
The factor $2$ in front accounts for the two WD-NS combinations due to the two NSs in the system.

\item \emph{GW inspirals:} The NS-NS GW inspiral cross section is found to be  
${\sigma}_{\rm insp,GW} \approx 1\cdot[1.2\cdot(5.1{\cdot}10^{-6})/10^2]{\cdot}2095{\cdot}(1075.0/5.1{\cdot}10^{-6})^{2/7} \text{AU}^2 \approx 0.03 {\rm AU}^2$,
by using Equation \eqref{eq:sigma_insp_GW}.

\item \emph{Collisions:} The WD-NS collision cross section is found from Equation \eqref{eq:sigma_coll_p} to 
be ${\sigma}_{\rm coll} \approx 2\cdot[1.2\cdot0.006/10^2]\cdot924\ \text{AU}^2 \approx 0.1\ \text{AU}^2$.

\end{itemize}

In this particular example we see that the inclusion of tides results in a total WD-NS coalescence cross section
that is about four times higher than the one estimated from the simple sticky star collision criterion.
One can compare these estimates with the upper left plot in Figure \ref{fig:cs_all}.
In an upcoming paper we extend this analytical framework to systems
where the WD can have any mass.

\subsubsection{Summary: Analytical Estimation of Cross Sections}

\begin{figure*}
\centering
\includegraphics[width=\textwidth]{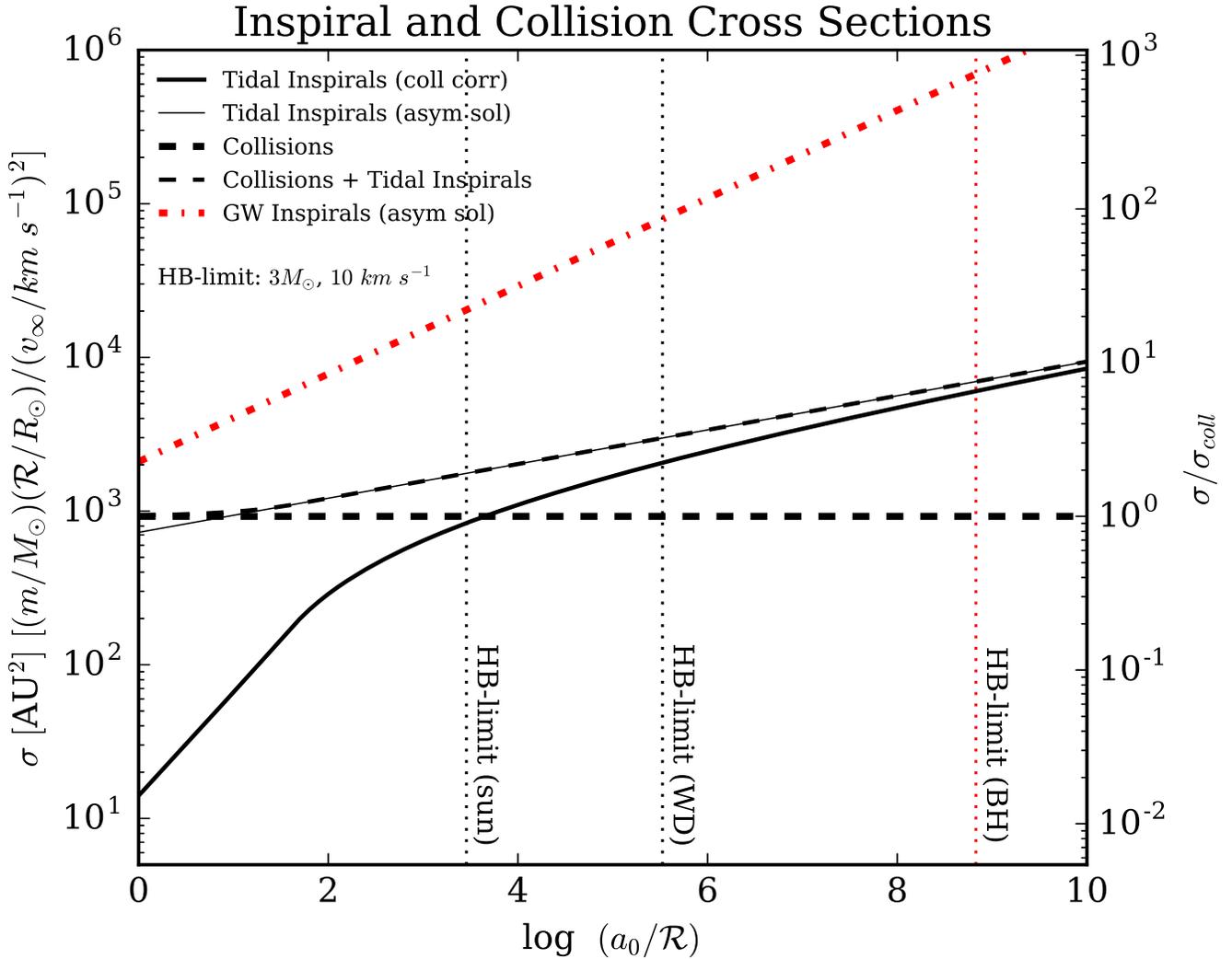}
\caption{Summary of outcome cross sections $\sigma$ related to finite size effects (collisions), tidal effects (tidal inspirals) and GR effects (GW inspirals)
arising from equal mass binary-single interactions. The black lines show the cross sections for an extended tidal object and a point-mass perturber to either
undergo a tidal inspiral (thick black solid line) or a sticky star collision (thick horizontal dashed line). The thin black line illustrates the analytical asymptotic solution
to the tidal inspiral cross section where the thin dashed line shows the collision + tidal inspiral cross section. The thick black line deviates from
the asymptotic solution due to the overlap between collisions and tidal inspirals in orbital phase space at especially small $a_{0}/R$ (see discussion
in Section \ref{sec:Inspirals in the collision dominated limit}). The tidal inspiral cross section shown here is valid for $n=3$ polytropes.
The red dash-dotted line shows the cross section for two compact objects (NS or BH) to undergo a GW inspiral. 
The left y-axis shows $\sigma$ in units of $[(m/M_{\odot})(\mathcal{R}/R_{\odot})/(v_{\infty}/\text{km}\ \text{s}^{-1})^{2}]$, where $v_{\infty}$ is the velocity
dispersion, $m$ is the mass of one of the (equal mass) interacting objects, and
$\mathcal{R}$ is the corresponding radius, which for tidal inspirals and collisions is the physical radius and for GW inspirals is
the Schwarzschild radius. A few examples are given in Section \ref{sec:Numerical Calibration of Analytical Cross Sections}.
The right y-axis shows $\sigma$ in units of $\sigma_{\rm coll}$. 
The functional form of the cross sections are based on our analytical
framework from Section \ref{sec:Analytical Models}, where the normalizations are estimated using our simulations with tides and
GR from Section \ref{sec: Numerical Scattering Results}. 
The three vertical dotted lines show from left to right $a_{\rm HB}/R$ for a system with $v_{\infty} = 10\ \text{km}\ \text{s}^{-1}$ and
$m = 1M_{\odot}$ (corresponding to $a_{\rm HB}\approx 13.3 \text{AU}$), when $R=1R_{\odot}$ (solar type star),
$R=0.0086R_{\odot}$ (WD) and $R=4.24{\cdot}10^{-6}$ (BH Schwarzschild radius), respectively.
The corresponding cross sections are only valid to the left of these lines. At this velocity dispersion, tidal inspirals with a solar type star can not
dominate the coalescence rate within the hard HB limit, but if the tidal object is a WD then tides can actually lead to an enhanced
coalescence rate by about a factor of four. The cross sections from this plot are overplotted our simulation data in Figure \ref{fig:cs_all}.
}
\label{fig:crosssec_summfig}
\end{figure*}

The calibrated cross sections from the section above including collision corrected tidal inspirals
are plotted and discussed in Figure \ref{fig:crosssec_summfig}. 
The tidal and GW inspiral cross sections are also shown in Figure \ref{fig:cs_all} with dashed and dashed-dotted lines, respectively. We see that our derived scalings do indeed work
all the way from a WD to a MS star across almost four orders of magnitudes in $a_{0}$.
Our analytical predictions give valuable insight into how the collisions and inspirals
possibly scale around the HB limit.

\section{Discussion}\label{sec:Discussion}

Our main findings, their consequences and relative importance in different dynamical systems
are discussed below.

\subsection{Stellar Collision Rate Not Enhanced By Tides}

Our initial motivation for this study was to explore if the modified dynamics
arising from tidal modes coupling to the orbital motion can enhance mergers or collisions 
in chaotic binary-single interactions. Using full numerical simulations and analytical arguments, we have learned
that the most significant change when including tides is the
formation of tidal inspirals -- similar to tidal captures in the field.

The main reason why the collision  or merger rate is not drastically 
altered by tides is that this would require the resonant system to undergo (at least) two independent close passages;
one that first drains some of the orbital energy through tides without leading to an inspiral, and then one that results in the actual merger.
However, mergers -- and thereby close passages -- are relatively rare, therefore the rate of collisions  following a previous close passage will happen only rarely. If the collision
probability is $P_{\rm coll}$ then the tidally induced collision probability will be of order $\approx P_{\rm coll}^2$. For example, in the WD case from Section \ref{sec:WDNSNS},
$P_{\rm coll} \approx 10^{-3}$ at $a_{0} = 1$AU. In fact, tides tend instead to decrease the number of collisions because
a system that could have evolved into a collision now can end as an inspiral. This is only seen at very low $a_{0}$.

We initially speculated that if tides could turn a fraction of the DIs into RIs (the encounter could be tidally captured into the triple system),
 the collision probability will be enhanced. However, the effective cross section for this to happen is simply too small. If it did happen,
the maximum enhancement would still only be about a factor of two since the ratio between the number of DIs and RIs initially is
about unity in the equal mass case \citep{2014ApJ...784...71S}.
A barrier for forming actual mergers is also the angular momentum, $L$. Even our inspirals, which represent the
highest energy and momentum loss configurations in our simulations, do not collide due to the
requirement of $L$ to be (almost) conserved (see Section \ref{sec:Tidal Capture Example}).
For a more accurate description one must include mass loss and dissipation, which requires the use of hydrodynamical simulations \citep{2010MNRAS.402..105G}.
Further discussion on the collision rates in general N-body systems can be found in \citet{2012MNRAS.425.2369L,2015MNRAS.450.1724L}.

\subsection{Tidal Inspirals and Collisions in Cluster Systems}\label{sec:Inspiral Mergers Relative to Collisions}
Studies indicate that tidal captures
are more likely to merge than to form a stable binary; see e.g. discussion in \citet{1993PASP..105..973R}.
A key question is then at what fraction our 3-body tidal inspiral mergers contribute to the
stellar coalescence rate compared to the classical sticky star collisions.
We can use our analytical framework to gain some insight into this question
by considering the ratio between the tidal inspiral and collision cross sections given
by Equations \eqref{eq:sigma_insp} and \eqref{eq:sigma_coll}, respectively,
\begin{equation}
\frac{\sigma_{\rm insp}}{\sigma_{\rm coll}} \propto \left( \frac{a_{0}}{R} \right)^{1/\beta}.
\label{eq:sigma_insp_sigma_coll}
\end{equation}
This relation shows that the rate of inspirals relative to collisions
increases as the size of the interacting objects decreases
and as $a_0$ increases.
The maximum value of $\sigma_{\rm insp}/{\sigma_{\rm coll}}$
for an object with radius $R$, is set by the hard binary limit
$a_{\rm HB} \propto m/v_{\infty}^{2}$ (see Equation \ref{eq:v_c}), from which we derive
\begin{equation}
\text{max}\left( \frac{\sigma_{\rm insp}}{\sigma_{\rm coll}} \right) \propto  \left( \frac{1}{v_{\infty}^2}\frac{m}{R}\right)^{1/\beta} \propto \left( \frac{v_{\rm esc}}{v_{\infty}}\right)^{2/\beta},
\label{eq:sigma_insp_sigma_coll}
\end{equation}
where $v_{\rm esc}$ is the escape velocity of the tidal object. 
From the equation above we can  conclude that the more compact the
interacting objects are (i.e., the larger $m/R$ is), 
the more inspirals can form relative to collisions.
This is also seen in our simulation results described in Section \ref{sec: Numerical Scattering Results},
and in Figure \ref{fig:crosssec_summfig}.

\begin{figure}
\centering
\includegraphics[width=\columnwidth]{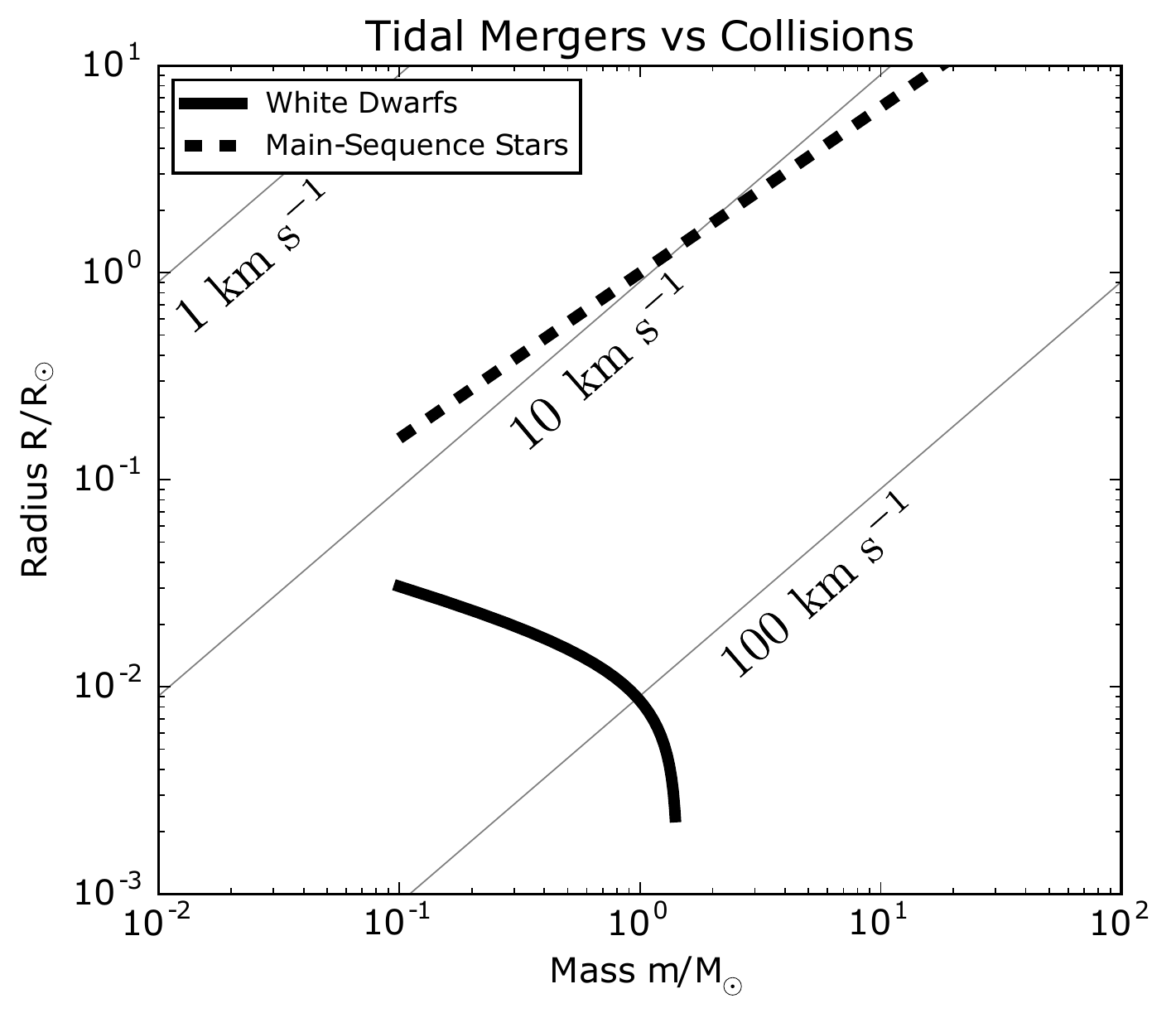}
\caption{Relation between compactness $m/R$, velocity dispersion $v_{\infty}$, and the rate of tidal inspirals relative to collisions.
The thin solid grey lines show the combinations of $m$, $R$ and $v_{\infty}$ from Equation \eqref{eq:mR_coll_eq_insp}
that will result in an equal number of tidal inspirals and collisions in the HB limit ($a_{0} = a_{HB}$). We assume the equal mass
case and the tidal inspirals are here between an extended tidal object and a point-mass perturber.
Also shown  are simplified mass-radius relationships for white dwarfs (solid line) and main sequence stars (dashed line).
If a given combination of $m$ and $R$ is to the right of a grey line, tidal inspirals will dominate over collisions in the HB limit for the
corresponding $v_{\infty}$. We see that  WDs are the only objects compact enough to produce a significant number
of tidal inspirals relative to collisions in a typical GC ($10\ \text{km}\ \text{s}^{-1}$), where MS star tidal inspirals probably only contribute to the coalescence rate in
open clusters ($1\ \text{km}\ \text{s}^{-1} $).
}
\label{fig:mR  Summery plot}
\end{figure}

The compactness $m/R$ required for an object to produce tidal inspirals with a point-mass perturber
at the same rate as collisions, can be read off Figure \ref{fig:crosssec_summfig}. This figure shows
that the tidal inspiral rate is similar to the collision rate when $\text{log}(a_{0}/R) \approx 3.5$.
If we use  the hard binary value $a_{\rm HB}$ from
Equation \eqref{eq:v_c}, we find 
\begin{equation}
\frac{m/M_{\odot}}{R/R_{\odot}} \approx 10^{-2} \left(\frac{v_{\infty}}{\text{km}\ \text{s}^{-1}}\right)^2 \;{\rm when}\; \sigma_{\rm coll} \approx \sigma_{\rm insp}(a_{\rm HB}).
\label{eq:mR_coll_eq_insp}
\end{equation}
This gives the critical value of $m/R$ in the HB limit. That is, if the tidal object has an $m/R$ larger than this value
then tidal inspirals can dominate over collisions.
Figure \ref{fig:mR  Summery plot} shows the relation from Equation \eqref{eq:mR_coll_eq_insp}
for different values of $v_{\infty}$, together with some simplified mass-radius relations for MSs ($R \propto M^{0.8}$ -- dashed line) and
WDs (see \cite{Zalamea:2010eu} -- solid line). 
We see that if the tidal object is a WD, tidal inspirals can  be as important as collisions
in GC systems ($v_{\infty}=10\ \text{km}\ \text{s}^{-1}$) and might even also play a role in
galactic nuclei ($v_{\infty}=100\ \text{km}\ \text{s}^{-1}$).
If the tidal object is a MS star, the rate of inspirals is much
lower compared to the rate of collisions, and inspirals will only contribute to the coalescence
rate in clusters with a $\approx 1\ \text{km}\ \text{s}^{-1} $ dispersion.
Interestingly enough, low dispersion clusters do have a high fraction of
wide binaries and are also surprisingly dynamically active \cite[see discussion in e.g.][]{2013MNRAS.432.2474L}.
Again, to make this picture applicable for describing more realistic astrophysical scenarios we need to
carefully work out the unequal mass case. 

\subsection{GW and Electromagnetic  Signatures from Tidal Inspirals}
\begin{figure}[tbp]
\centering
\includegraphics[width=\columnwidth]{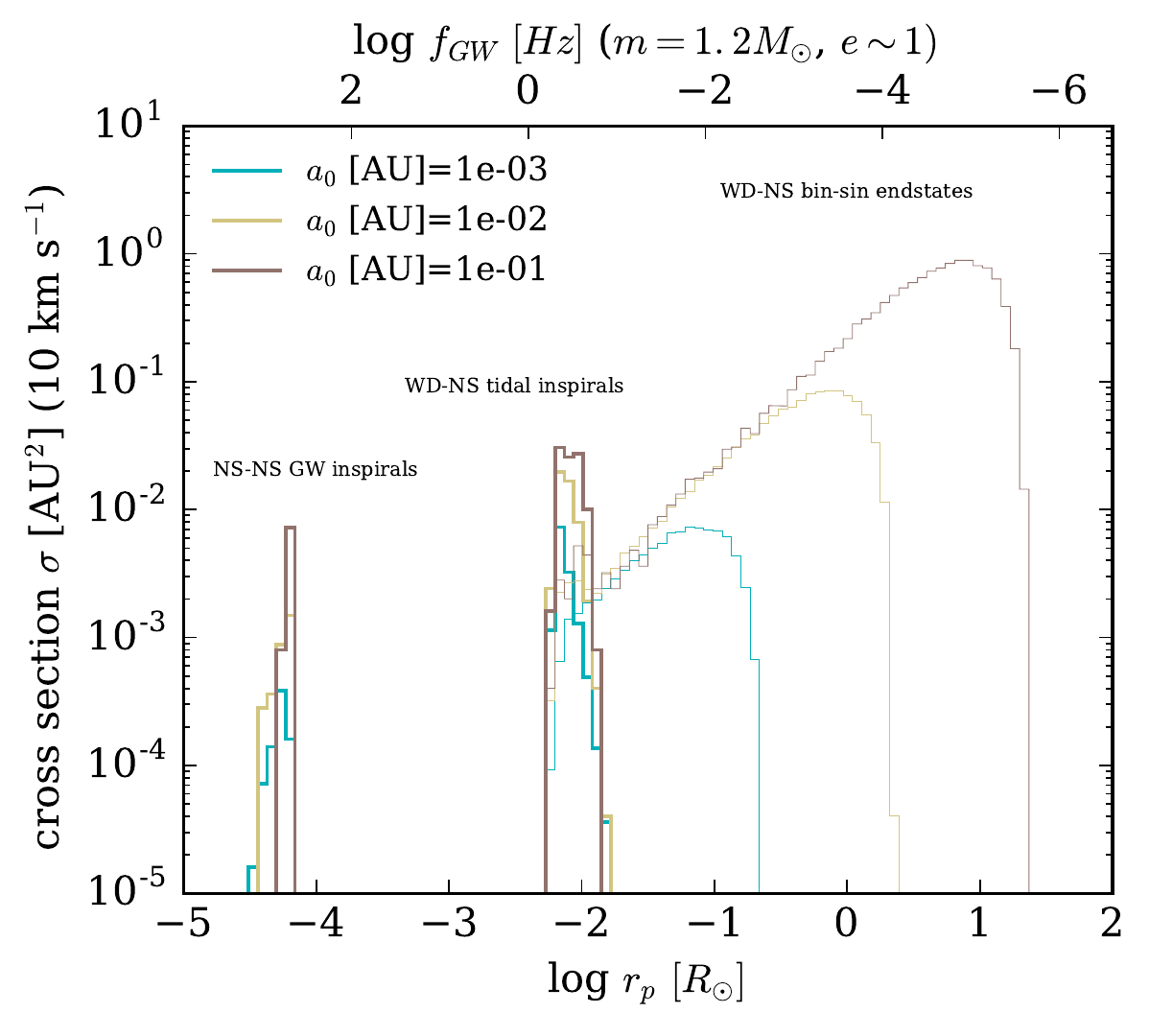}
\caption{Outcome cross section $\sigma$, as a function of the pericenter distance $r_{\rm p}$ of the endstate binary,
computed from a set of $5\times10^{4}$ binary-single interactions between a [NS($1.2M_{\odot}$), WD($1.2M_{\odot}$)] binary and a single
incoming NS($1.2M_{\odot}$). 
The \emph{thin solid lines} show the distribution of WD-NS binaries from binary-unbound-single endstates (such as an exchange or a fly-by).
The \emph{thick solid lines} show the distribution of inspirals, where the left and right peak
is the NS-NS GW inspirals and WD-NS tidal inspirals, respectively.
These inspiral populations only appear when tides and GR are included in the EOM of the N-body system.
Inspirals are characterized by very low angular momentum
corresponding to a small pericenter when $e\sim1$, which makes them interesting sources for both EM and GW signals.
From the upper axis showing the corresponding GW frequency $f_{\rm GW}$, we see that GW inspirals fall within the LIGO sensitivity band, where
the tidal inspirals are closer to the LISA band. These results greatly motivates further studies of compact objects
undergoing a high eccentricity evolution.  The cross section as a function of $a_{0}$ is shown for the same set in Figure \ref{fig:cs_all}.
}
\label{fig:L2m}
\end{figure}

Tidal and GW inspirals are characterized by high eccentricity and low angular momentum (Figure \ref{fig:L2m}).  The high eccentricity especially allows for multiple close passages before merger which will give rise to unique electromagnetic (EM) and GW observables, especially when the tidal object is
a WD \citep{2009PhRvD..80b4006P, 2011PhRvD..83f4002P, 2011PhRvD..84j4032P}. The GW signal will have a very rich spectrum compared to normal circular inspirals \citep{2007ApJ...665L..59W, 2010MNRAS.406.2749L},
which will reveal much more information about especially the equation of state of the WD.

As seen in Figure \ref{fig:L2m}, a space-borne GW
instrument like LISA will be sensitive to these WD-NS inspirals.
However, while there are plenty of interesting physics in high eccentricity WD-NS tidal inspirals and collisions, the rates are expected to be modest from the binary-single channel: 
if we consider the $0.6M_{\odot}$ WD case from \cite{2014ApJ...784...71S} and assume that the inclusion of tides
enhances the resultant merger rate by a factor of $5$ (a $0.6M_{\odot}$ WD both has a lower polytropic
index $n$ and an $\alpha \sim 0$, which is expected to lead to more inspirals compared to a heavy WD),
then the expected rate of WD-NS tidal
inspirals will be around $\approx 50\ \text{yr}^{-1} \text{Gpc}^{-3}$.
The problem here is that the associated GW strain is far too weak for these sources to been seen
outside our own galaxy by LISA.  More promising  signatures could be thermonuclear optical transients
\citep{1986SvAL...12..152K, Lee:2007em, 2009MNRAS.399L.156R, Rosswog:2009wq, 2010ApJ...714L..52S,
2010Natur.465..322P,  2011ApJ...738...21W, 2012ApJ...746...62R, Metzger:2012ge, 2013ApJ...771...14H}, and high-energy transients 
\citep{1998ApJ...502L...9F, 1999ApJ...520..650F, Lee:2007em}
that are expected to ensue when  both light and heavy WDs are shocked in collisions or mergers with COs. 
The exact rates of such encounters requires a detailed understanding of unequal mass scatterings involving WDs and COs with tides and GR, which we plan 
to consider in future work.

\section{Conclusion}\label{sec:Conclusion}

We present the first systematic study of how dynamical tides affect the
interaction and relative outcomes in binary-single interactions. From performing a large set
of binary-single scatterings using an $N$-body code that includes tides and GR, we find that the inclusion of tides
leads to a population of tidal captures which are occurring during the chaotic
evolution of the triple interaction. We denote these captures \emph{tidal inspirals}, partly due to their similarity with
the GW inspirals studied in \citep{2014ApJ...784...71S}.

We confirm with analytical models
that the rate of tidal inspirals relative to the classical sticky star collision rate
increases with $(a_{0}/R)$, as a result, tides show the largest effect for widely separated binaries.
Since the upper limit on $a_{0}$
is set by the HB limit, which scales linearly with mass $m$, we conclude that the
compactness $m/R$ of the tidal object is the key factor
for determining if tides play a significant role or not in a given cluster environment: a larger compactness leads to more
tidal inspirals relative to collisions. As a result of these scalings, we find that the only tidal object which is compact enough to have
tidal inspirals dominating over collisions in a typical GC environment is a WD.

We further conclude that tides, from a dynamical perspective, do not seem to effect the dense stellar system as a whole,
as otherwise speculated in several previous studies \citep[e.g.][]{Fregeau:2004fj} --  although stellar finite sizes do
matter through collisions and dynamical kicks \citep{1986ApJ...306..552M}. However, the inclusion
of tides and GWs leads to a rare, but highly interesting population of eccentric binaries. The high eccentricity likely results in unique EM and GW
signals.  While highly eccentric binaries can be created in single-single captures, it was illustrated
in \cite{2014ApJ...784...71S} that the binary-single channel is likely the dominant
formation path. These observations motivate further dynamical studies on few-body interactions involving
especially WDs and COs, as well as hydrodynamical studies on the outcome of highly eccentric captures. 

While our estimated inspiral rate involving a heavy WD (1.2$M_{\odot}$) is still modest, we do expect the rate to be
significantly higher for lower mass WDs simply because they are more vulnerable to tidal deformations. 
We are currently working on the analytical prescriptions for unequal mass encounters.

\acknowledgments{
It is a pleasure to thank D. Spergel, R. Cen, V. Paschalidis, C. Holcomb, T. Ilan, and F. Pretorius
for helpful discussions. Support for this work was provided by  the David and Lucile Packard Foundation, UCMEXUS (CN-12-578) and  NASA through an Einstein
Postdoctoral Fellowship grant number PF4-150127, awarded
by the Chandra X-ray Center, which is operated by the
Smithsonian Astrophysical Observatory for NASA under
contract NAS8-03060.
}

\bibliographystyle{apj}


\end{document}